\documentclass[sigconf]{acmart}
\pdfoutput=1

\AtBeginDocument{%
  \providecommand\BibTeX{{%
    \normalfont B\kern-0.5em{\scshape i\kern-0.25em b}\kern-0.8em\TeX}}}
    
\setcopyright{none}

\renewcommand\footnotetextcopyrightpermission[1]{} 

\newcommand{\tc}{\emph{Time Constant}\xspace} 
\newcommand{\name}{{\em Time Constant}}
\usepackage{tikz}
\usepackage{algorithmic}
\usepackage{graphicx}
\usepackage{subcaption}
\usepackage{textcomp}
\usepackage{xcolor}

\usepackage{multicol}
\usepackage{multirow}
\usepackage{tabularx}

\usepackage{booktabs} 
\usepackage{adjustbox}
\usepackage{xspace}

\usepackage{soul}

\usepackage{listings}
\usepackage{xcolor}

\definecolor{codegreen}{rgb}{0,0.6,0}
\definecolor{codegray}{rgb}{0.5,0.5,0.5}
\definecolor{codepurple}{rgb}{0.58,0,0.82}
\definecolor{backcolour}{rgb}{0.95,0.95,0.92}
 
\lstdefinestyle{mystyle}{
    backgroundcolor=\color{backcolour},   
    commentstyle=\color{codegreen},
    keywordstyle=\color{magenta},
    numberstyle=\tiny\color{codegray},
    stringstyle=\color{codepurple},
    basicstyle=\ttfamily\footnotesize,
    breakatwhitespace=false,         
    breaklines=true,                 
    captionpos=b,                    
    keepspaces=true,                 
    numbers=left,                    
    numbersep=5pt,                  
    showspaces=false,                
    showstringspaces=false,
    showtabs=false,                  
    tabsize=2
}

\begin{document}

\title{Time Constant: Actuator Fingerprinting using Transient Response of Device and Process in ICS}






\author{Chuadhry Mujeeb Ahmed}
\affiliation{%
  \institution{Newcastle University}
  \city{Newcastle upon Tyne}
  \country{United Kingdom}}
\email{mujeeb.ahmed@newcastle.ac.uk}

\author{Matthew Calder}
\affiliation{%
 \institution{University of Strathclyde}
 \city{Glasgow}
 \country{United Kingdom}}
\email{matthew.calder@strath.ac.uk}

\author{Sean Gunawan}
\affiliation{%
  \institution{Singapore University of Technology and Design}
  \city{Singapore}
  \country{Singapore}}
\email{sean_gunawan@sutd.edu.sg}

\author{Jay Prakash}
\affiliation{%
  \institution{Silence Labs}
  \city{Singapore}
  \country{Singapore}}
\email{jay_prakash@mymail.edu.sg}

\author{Shishir Nagaraja}
\affiliation{%
  \institution{Newcastle University}
  \city{Newcastle upon Tyne}
  \country{United Kingdom}}
\email{shishir.nagaraja@newcastle.ac.uk}

\author{Jianying Zhou}
\affiliation{%
  \institution{Singapore University of Technology and Design}
  \city{Singapore}
  \country{Singapore}}
\email{jianying_zhou@sutd.edu.sg}

%


\begin{abstract}
  Command injection and replay attacks are key threats in Cyber Physical Systems (CPS). We develop a novel actuator fingerprinting technique named {\name}. {\name} captures the transient dynamics of an actuator and physical process. The transient behavior is device-specific. We combine process and device transient characteristics to develop a copy-resistant actuator fingerprint that resists command injection and replay attacks in the face of insider adversaries. We validated the proposed scheme on data from a real water treatment testbed, as well as through real-time attack detection in the live plant. 
  Our results show that we can uniquely distinguish between process states and actuators based on their {\name}. 
  \keywords{CPS/IIoT  \and Sensor-Actuator Security \and Fingerprinting.}
\end{abstract}

\maketitle
\pagestyle{plain} 

\section{Introduction}
\label{sec:Introduction}
Industrial Control Systems~(ICS) control modern critical infrastructures, including but not limited to smart grids, water treatment plants, and digital manufacturing units. Connectivity in an ICS provides improved monitoring and operation of a physical process but makes it challenging to secure modern ICS~\cite{cardenas2009challenges,Challenges_SmartGrid_del}. A typical ICS has three major physical subsystems, i.e., sensors, processes and actuators. Each of these components comes with its vulnerabilities and threats. A lot of attention has been given to attacks on sensors and corresponding defense technologies~\cite{shoukry2015,yasser-2013,drone_Son2015,sensor_saturationAttack_infusionpump_usenix2016,sampling_race2016,walnut_acoustic_attack_mems_accelerometer,urbina_CCS2016limiting,Survey_Jairo_Cardenas_2018,panoff2021sensor,Yan2016CanYT_WenyuanXu_Defcon2016,ICS_zhejiany_wenyuan,WenyuanAndCo}, but little attention is paid to the actuator and process security~\cite{raheem2016,actuationAttack_asiaccs2018_RyanGerdes,shekari2019rfdids_Tohid_NDSS2019,actuator_saturation,actuator_attacks_UAV_del}. Actuator subsystems are a fundamental part of the process industry wherein each actuator in itself is a dynamic system \cite{zhang2018root_cause_analysis_ActuatorFault}. An attack on actuators may cause serious accidents as in the case of Stuxnet~\cite{stuxnet}, reduction in product quality and levels of contamination as happened in the 2021 attack on the water treatment system of Oldsmar, Florida~\cite{Florida_PoisoningAttack_Feb2021}. 

Command injection and replay attacks have been observed in the wild. A few examples include Stuxnet malware and Ukrainian power grid shutdown~\cite{ukranian_case2016analysis}. In both cases, Supervisory Control and Data Acquisition (SCADA) system, as well as the actuator subsystem, were attacked. In the case of Stuxnet, the malware modified the variable frequency drive controlling centrifuge speed by changing the frequencies between 1410 Hz and 2 Hz, to disrupt the Uranium enrichment process. However, normal data was reported to the operators since the SCADA system was under a replay attack. Such a sophisticated attack could not be detected by network monitoring technologies alone as there is no mechanism to validate the control commands~\cite{SRID_invariants_ESORICS2014}. In the case of the Ukrainian power grid attack, the breakers in 30 substations were switched off and the SCADA network traffic was spoofed to appear normal to the control center~\cite{shekari2019rfdids_Tohid_NDSS2019}. 
Recently, a live-fire cyber attack-defense exercise on an ICS, demonstrated that the network-only solutions do not succeed in detecting process layer attacks~\cite{iTrustCISS2017Report_NetworkMonitoringNotEnough}.

A Programmable Logic Controller (PLC) controls the actuators based on the sensor measurement. If an adversary can spoof actuator data in the digital or physical domain, it can drive the system to an unsafe state~\cite{NIPAD_powerbased_detection_PLC,StateEstimator_Del}. 
An attack on an ICS would not only damage the IT part of it but also the physical domain~\cite{okhravi2009application_del}, resulting in catastrophic outcomes in the physical space and on the lives of people~\cite{aurora_attack,stuxnet,ukranian_case2016analysis}. The actuator subsystem can directly affect the physical space, therefore, it is important to validate the physical state reported to the controller and authenticate the actuator. In this work we propose to use the time that an actuator takes to operate (e.g., open/close or ON/OFF operation) as its fingerprint. We propose two techniques, one passive called \tc and one active called \emph{Control Signal Watermarking}.

\noindent \tc: We model a relationship between the operation of an actuator to the output channel i.e., an associated sensor. This means that we can extract the timing information by analyzing a sensor's data rather than the control command timings, directly. The second aspect is the transient timing characteristics of a process to make it even harder for an attacker to forge the readings. For example, in a water pumping system, when a pump turns ON it takes a finite amount of time for the water flow rate to reach the maximum value. This is defined as a process transient. A combined fingerprint for device motion timings and process transients timing is developed and called herein as the \tc. 
\noindent \emph{Control Signal Watermarking}: To tackle the attackers who can learn and add the \tc pattern in the attacked sensor measurements, we propose a watermarking technique. Since the fingerprint is in units of time, a random time delay is injected in the \tc through the control action and it is recovered in the sensor measurements. \textbf{Contributions}: In this paper we proposed a novel technique to fingerprint the actuators and the physical state of an actuator.  The main contributions of this work are, 1) To design and test a process transient based actuator fingerprinting technique. 2) To design a practical watermarking technique to extend \tc and defend against powerful attacks, e.g., replay attacks.

\section{Ideation and Threat Model}
\label{Sec:Background}

\begin{figure}
    \centering
    \includegraphics[scale=0.5]{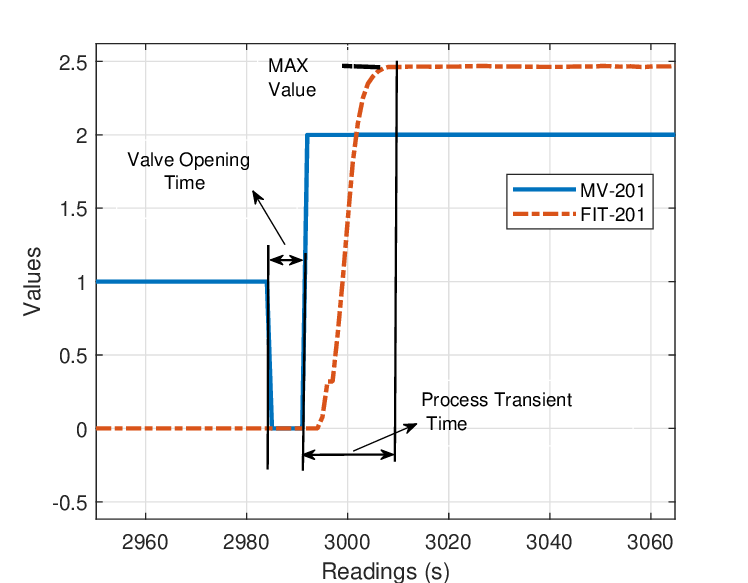}
        \caption{The idea: Fingerprint is based on the temporal characteristics of device movement and process transients. }
    \label{fig:the_idea}
\end{figure}


\subsection{Idea and Intuition}
Industrial Control Systems have a broad domain. In this work, a Secure Water Treatment testbed~(SWaT)~\cite{swat2016} is used primarily which is pictorially shown in Figure~\ref{fig:swat_pic} in Appendix \ref{appendix_supporting_table}. Figure~\ref{fig:the_idea} provides a nice example to demonstrate the intuition behind our technique. Two signals are plotted in Figure~\ref{fig:the_idea}, namely motorized-valve control action in stage~2 of the SWaT testbed labeled as MV-201 and flow sensor measurement labeled FIT-201. The flow of water depends on the control action of the motorized valve. When the motorized valve is closed (has a value of 1), the flow rate is zero. When the motorized valve is open (has a value of 2), water flows through the pipe at a maximum rate of $2.4 m^3/hr$. We observe that the motorized valve takes time to open, demonstrating transient behaviour which can be captured by downstream sensors. This is the property that lets us fingerprint a motorized valve by examining sensor measurements. 

To ensure safety, the control logic in the SWaT testbed turns on the pump only if the motorized valve is open. In other words, a pump and a motorized valve are logically interlocked. This can be seen in Figure~\ref{fig:the_idea}, where the water starts flowing in the pipe only after the MV is fully open, i.e., it is in state 2. We observe that transient behaviour of MV such as the time taken to open fully is captured by flow sensor measurements, given the status of the control command. The second important observation from Figure~\ref{fig:the_idea} is that when water starts flowing through the pipe it takes time to achieve the full flow rate. This time from zero to full flow rate is called the transient time of the water flow process. The transient behavior exists due to the resistance to the flow of the water and due to the dynamics of the process, that is, a change in a process would not happen in zero time. The resistance in the fluid system is caused by friction between the outer layer of fluid and the walls of a pipe and friction between fluid layers~\cite{resistance_fluid}. This transient behavior is similar to the transient behavior of the charging circuit of an electric capacitor, hence the title \emph{Time Constant}. 
\vspace{-1mm}


\noindent  The process transients are due to a physical process and are a property of the associated process. Given constant pumping rate and the fixed size pipe, as is the case in a real industrial system, it is indeed possible to create a fingerprint for the process. It is difficult (but possible) for an attacker to replicate such transient behaviors of a real physical process. The device timing characteristics are specific to a device and we hypothesize that based on these timing characteristics even similar actuators can be fingerprinted due to the manufacturing imperfections of the components used in each device. In the example shown in Figure~\ref{fig:the_idea} it is possible to separate the device opening time and the process transient timings, but in some other cases, it might not be the case. For example, consider an inflow of water solely controlled through a motorized valve and where no pump is involved. In that case, as soon as the motorized valve starts opening the flow process transient behavior also begins as the water starts flowing. In that case, both the device movement timing fingerprint and the process transient could be overlapping and hard to separate. Therefore, we propose a joint fingerprint termed as \tc.   





\subsection{Threat Model}
\label{sec:threat_model}

We assume an active-insider attacker who has access to the control plane and can therefore modify or inject control commands to the desired value. \tc based actuator fingerprinting can detect such malicious manipulation of the control signal during the state transition. Device fingerprinting techniques mine patterns from side channels to authenticate a device, however, a typical limitation of such fingerprinting techniques is a replay attack because the fingerprinting patterns get preserved under such an attack~\cite{raheem2016,Qinchn_Raheem_TDSC2021}. We consider replay attacks in this work as a worst-case scenario and extend the \tc with a watermark technique introduced as a random delay, to tackle such an attacker. An attacker can observe the delay on the network but is limited by its ability to react fast enough~\cite{shoukry2015} to the random watermark. We do not consider an attacker with a higher sampling rate~\cite{sampling_race2016} than the defender.  

\noindent \textbf{Security objectives}: 1) Verify whether PLC commands have been executed by actuator components using side-channel information. 2) Authenticate actuator side-channel information provenance to a legitimate actuator.

\begin{figure}
    \centering
    \begin{subfigure}{0.35\textwidth}
        \centering
        \includegraphics[width=\textwidth]{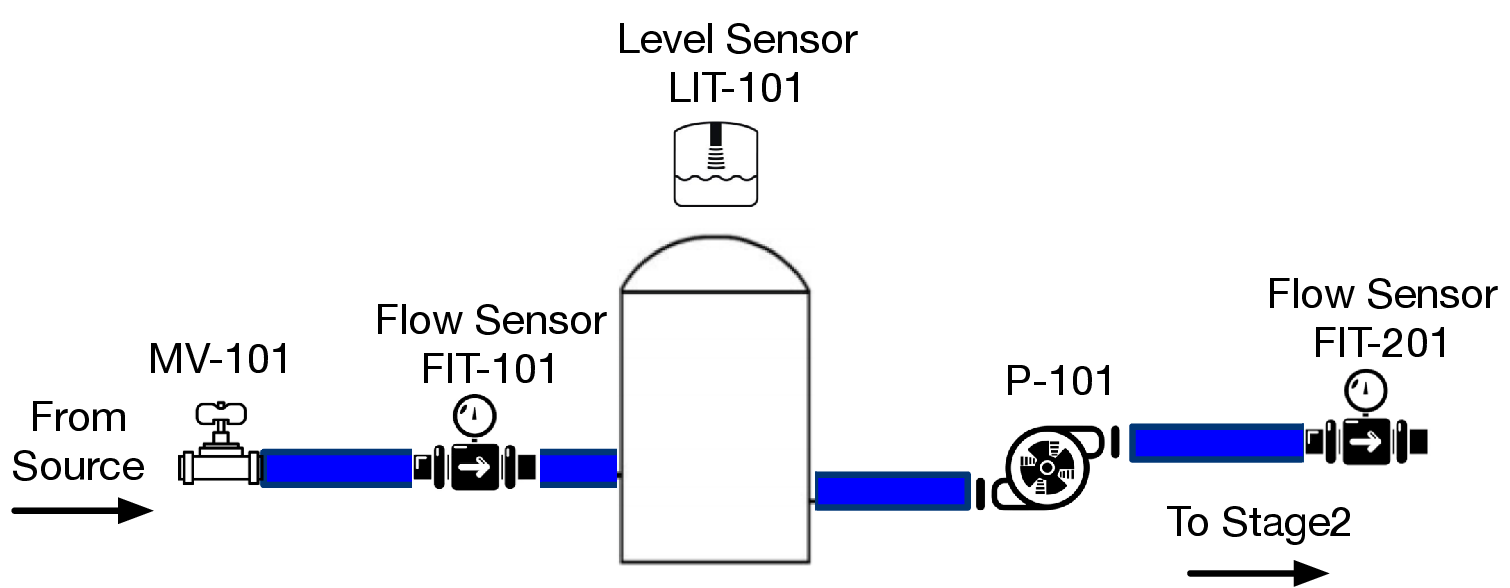}
        \caption{}
        \label{fig:stage1_diagram}
    \end{subfigure}
    \hfill
    \begin{subfigure}{0.35\textwidth}
        \centering
        \includegraphics[width=\textwidth]{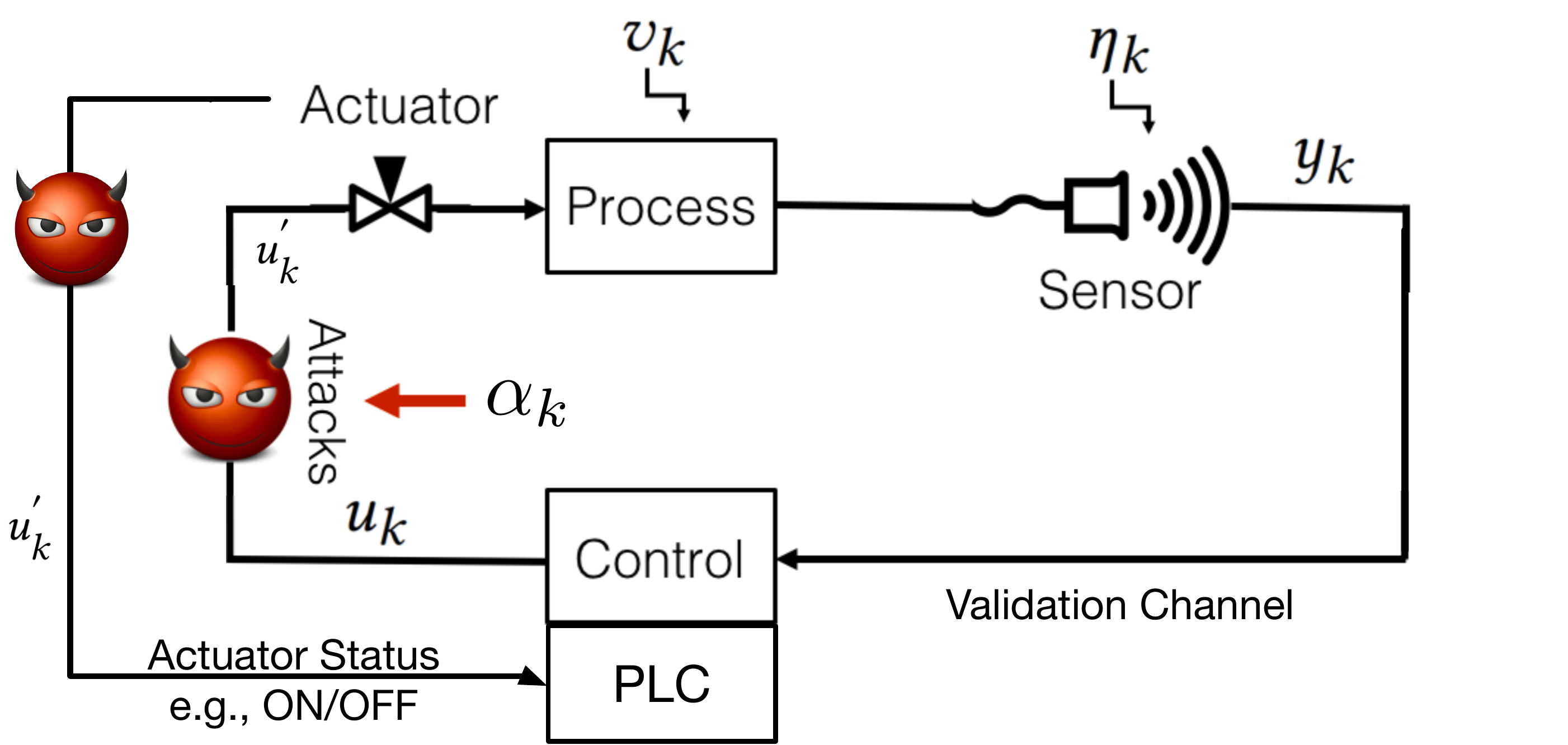}
        \caption{}
        \label{fig:attack_sensor_actuator}
    \end{subfigure}
    \caption{(a) Stage~1 of SWaT as a running example. MV:motorized valve, FIT: flow sensor, LIT: level sensor and P:pump. (b) Attack on the control command channel between a PLC and an actuator. Control commands and actuator responses are vulnerable to MITM, motivating the need to validate control plane.}
\end{figure}

\noindent \textbf{Justification}: An adversary can spoof actuator commands either through the cyber domain while data is being transmitted from/to controller/devices or in the physical plane. It has been shown in the recent studies~\cite{farraj2017impact_Attack_On_Actuator_Transients,kandasamy2019investigation_synchronization_attack} that injecting control commands during the transient dynamics are hard to detect and would lead to huge losses. In~\cite{kandasamy2019investigation_synchronization_attack} researchers have demonstrated that an attack on a real-world system where a small change in control leads to the generator synchronization process delayed by 1.5 hours, a process that normally takes a few seconds when a generator powers ON.  It is important to authenticate whether the data is originating from the real physical process or being altered. Due to computational limitations and legacy compliant equipment, it is not feasible to rely on cryptographic methods~\cite{john_acns2017_castellanos2017legacy}.
Figure~\ref{fig:attack_sensor_actuator} shows that an attacker can modify a rightful control command by a value $\alpha_k$. Figure~\ref{fig:attack_sensor_actuator} shows a generic closed-loop feedback control system. A PLC executes a control command to drive the physical process as required. The control command could be modified in transit, e.g., during a replay attack. Another signal from the actuator travels back to the controller to report the current state of the actuator, this can be considered an acknowledgment sent back to PLC to confirm that a particular control has been executed. However, this feedback command could also be spoofed to deceive PLC regarding the real status of the actuator. The closely related work in~\cite{raheem2016} would not succeed under such an attack, because they measure the operation timing of an actuator over this feedback channel itself. Moreover, the experiments conducted on only two latching relays in \cite{raheem2016}, used the application layer time stamps which are transmitted in plain text and could be modified by the attacker as demonstrated recently by~\cite{Qinchn_Raheem_TDSC2021}. 



\begin{figure}
    \centering
            \includegraphics[scale=0.3]{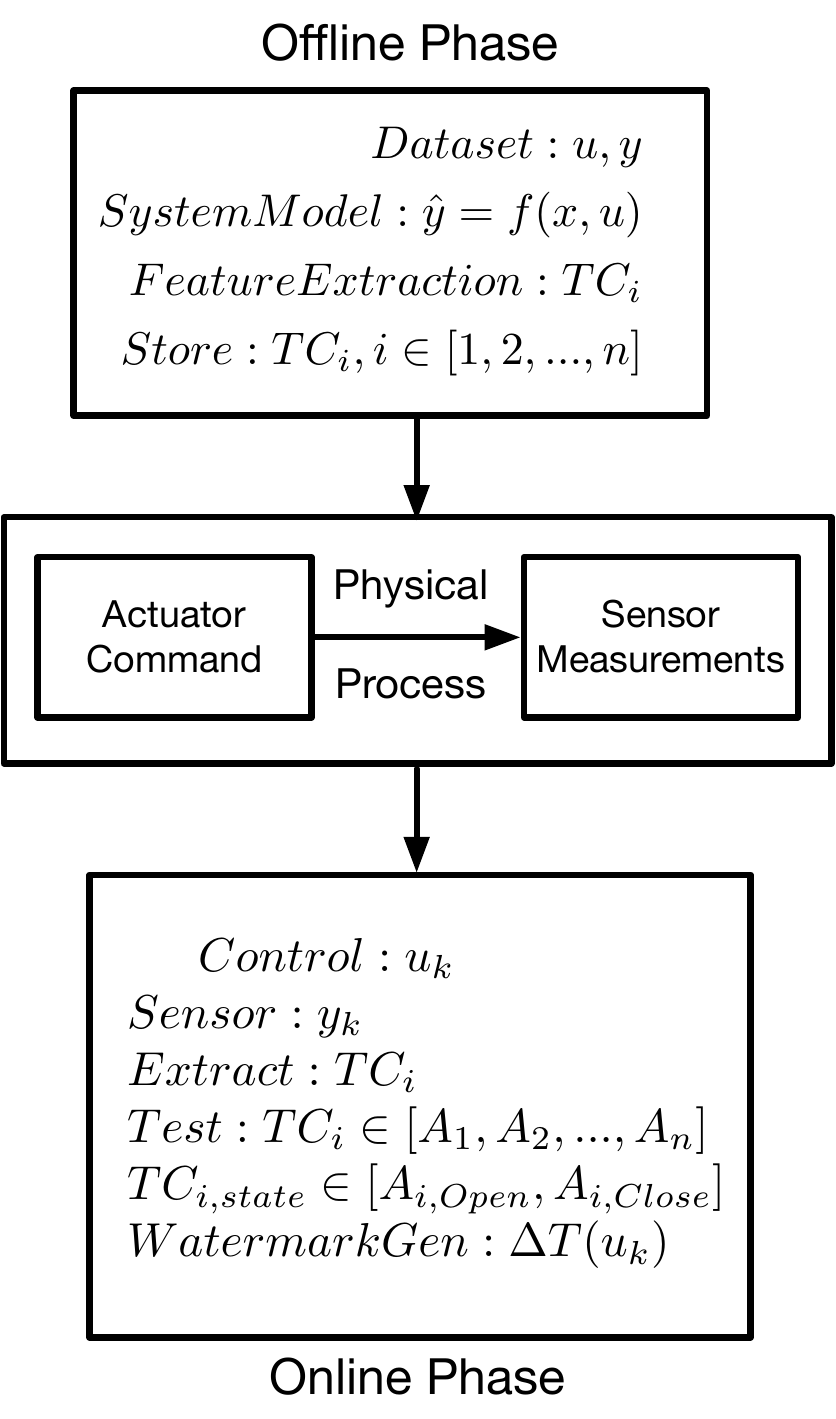}
        \caption{Design Overview: Offline phase extracts data composed of control commands and sensor outputs from the physical plane. Online phase computes \tc  fingerprints, whilst continually refreshing for actuators and processes profiles using fresh data.}
    \label{fig:tc_overview_diagram}
    
\end{figure}

\vspace{-3mm}

\section{Design of \tc}
\label{Sec:DesignOfTimeConstant}
\vspace{-1mm}
The core components of the design of \tc are laid out in Figure~\ref{fig:tc_overview_diagram}. There are two major phases, an offline phase, and an online phase. In the offline phase, the system model is obtained based on the data collected from an ICS. The data is composed of inputs (control commands: $u$) and outputs (sensor measurements: $y$). Based on the proposed idea of the \tc, the fingerprint is composed of the transient features (details to follow). Once the actuators are profiled, fresh sensor and actuator data can be obtained and provided as input to the online phase. The online phase takes similar steps to extract \tc fingerprints and test for the actuator identification as well as its state. A watermark signal is generated in real-time to tackle powerful attacks, e.g., replay attacks. All of these steps are explained in detail in the following but first some definitions to set the context.


\noindent \textbf{Definition} State of a Process: State of a process describes the system dynamics~\cite{palm2014system_dynamics_book}. In our case, it is the physical state of a system, for example, for a sensor the measured value by the sensor.



\noindent \textbf{Definition} Steady State: It is defined as the state of a process that has reached a stable value over time. 



\noindent \textbf{Definition} Transient State: It is defined as the state of a process before reaching the steady-state. 



\noindent \textbf{Definition} Sensor-Dependent Control: It is defined as a control action performed with respect to the values of a particular sensor.


\noindent \textbf{Definition} Sensor-Independent Control: It is defined as a control action performed irrespective of the values of a particular sensor. 

\begin{figure*}
    \centering
    \begin{subfigure}{0.6\textwidth}
        \centering
        \includegraphics[width=\textwidth]{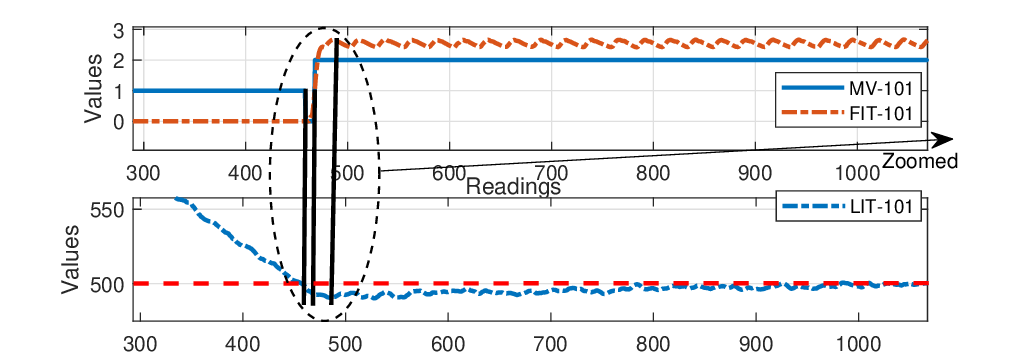}
        \caption{Process}
    \end{subfigure}
    \begin{subfigure}{0.33\textwidth}
        \centering
        \includegraphics[width=\textwidth]{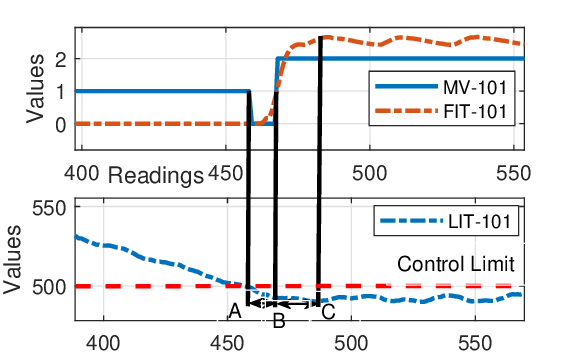}
        \caption{ZoomedIn}
    \end{subfigure}
    \caption{Closed-loop control (MV-101) based on level sensor (LIT-101). Flow sensor (FIT-101) directly captures the transients when MV-101 opens and LIT-101 reflect that transient due to the change in the water level from the low set-point.}
     \label{fig:dependence_sensor_actuator}
\end{figure*}



\vspace{-2mm}
\subsection{Motivating Example}
\noindent \emph{Sensor Dependent vs Independent Control}: Consider the stage~1 of the SWaT testbed as an example as shown in Figure~\ref{fig:stage1_diagram}. The control of the motorized valve (MV-101) depends on the level sensor (LIT-101). If LIT-101 reports a low level then PLC will send a command to open the inlet valve MV-101. On the other hand, the flow sensor (FIT-101) does not affect the control logic in the PLC. However, an interesting point is that both the flow and level sensors capture the behavior of the actuator, that is, the timing profile of the actuator as shown in Figure~\ref{fig:dependence_sensor_actuator}. Figure~\ref{fig:dependence_sensor_actuator} shows that the combination of both the opening time for the valve (time between point A and B in the figure) and transient time of the process (time between point B and C in the figure) together known as \tc is captured by both the sensors.  
In Figure~\ref{fig:dependence_sensor_actuator}, it can be seen that as soon as the water level (LIT-101) hits the low value of 500mm (lower set-point), the motorized valve (MV-101) starts opening, however, it takes some time to open and hence water level keeps dropping. When the flow at the inlet pipe reaches the steady state, only then do we see that the water outflow in the tank~1 stops and the filling process changes the direction of water level in the tank rising towards and above low set-point of 500mm. The time taken since hitting the 500mm low-level value until the change in the level direction upwards is quantitatively the same as \tc being measured on the flow sensor (FIT-101). Figure~\ref{fig:dependence_sensor_actuator} shows these measurements for both the sensors zoomed-in on the right-hand pane.  This brings us to the discussion of which sensor-actuator pair shall be used to measure the \tc. In the example above, it is observed that one actuator (MV-101) dynamics was captured by the two sensors, a flow sensor and a level sensor. Similarly, it is also possible to capture multiple actuator dynamics using one sensor as shown in Figure~\ref{fig:the_idea}. 
\vspace{-5mm}

\subsection{Sources of Fingerprints}
\label{sec:sources_of_fingerprint}
The automation in an ICS relies on a critical system known as an actuator-valve subsystem in general and is an essential part of ``smart ICS"~\cite{american2001butterfly_BearingTorque_2001}. The same system is taken as an example for the following discussion. A brief discussion on the sources of fingerprints is presented in 
the following.

\noindent \emph{Solenoid Actuator}: The valve-actuator systems studied in this work are industrial grade devices used in the real-world deployments. These devices commonly use solenoid coil~\cite{flutech_valve} arrangement to operate the valve. The electromagnetic force produced during the process of energizing the solenoid coil depends on the number of turns in the solenoid, cross-sectional area and core of the magnet. These are key parameters that induce device fingerprints due to the physical construction of the components and are known to be stable device fingerprints~\cite{raheem2016}.

\noindent \emph{Role of Fluid Dynamics}: We explore how the internal dynamics of a component be used for designing a defense technique. As an example, consider an electrically energized solenoid actuator (used in this work). The magnetic flux generates electromagnetic force to move the valve in an electro-mechanical construction. The valve controls the flow in a pipe and is hence subject to hydrodynamic forces \cite{ButterFly_Valve_Modelling_2011}. It is a well-known result that the opening and closing processes of the valve indicate differences regarding the behavior of hydrodynamic torque \cite{ButterFly_Valve_Modelling_2011}, endorsed by our observations as well and resulting in device state fingerprints. Everything else being constant, it is known that the torque acting on the valve depends on the process parameters, e.g., pipe diameter and inlet velocity. These along with other parameters constitute a unique fingerprint for the actuator and different operations, e.g., opening or closing.
\subsection{Process Dynamics}

\tc depends on the process dynamics and parameters as explained in the previous section. Physical processes are complex and it is hard to measure the effects of each parameter separately, therefore, we propose to capture the relationship among the different variables using the process dynamics modeled at a system level. 
In the form of Linear Time-Invariant (LTI) system of equations, it can be written as,\vspace{-3mm}

\begin{equation}
\left\{
\begin{array}{ll}
x_{k+1} = Ax_k + Bu_k + v_k, \label{system_model_eq}   \\
 y_k = Cx_k + \eta_k, 
\end{array}
\right.
\end{equation}

\noindent where $A,B$ are the state space and control matrices of appropriate dimensions. $x_k$ represents the system state, for example, the water level in a tank or water flow rate through a pipe at time step $k$. $v_k$ is the process noise. The physical system state defined by variable $x_k$ can be measured by using a sensor represented as $y_k$. The measurement matrix $C$ determines the measurable states. $\eta_k$ is the sensor measurement noise. 
We used a data-driven technique known as \emph{sub-space system identification}~\cite{van1996} to obtain a system model. 

\subsection{Feature Extraction}
\tc is measured for the actuators and collected as a vector over multiple operations of the device. We used the resultant pattern over number of runs to extract the time-domain features. We used the Fast Fourier Transform (FFT) algorithm\,\cite{welch1967} to convert data to the frequency domain and extract the spectral features. Empirically, a set of eight features (Table\,\ref{features} in Appendix~\ref{appendix_supporting_table}) is selected to construct the fingerprint. 



\subsection{Watermark Technique}





\begin{figure}
    \centering
        \includegraphics[width=0.35\textwidth]{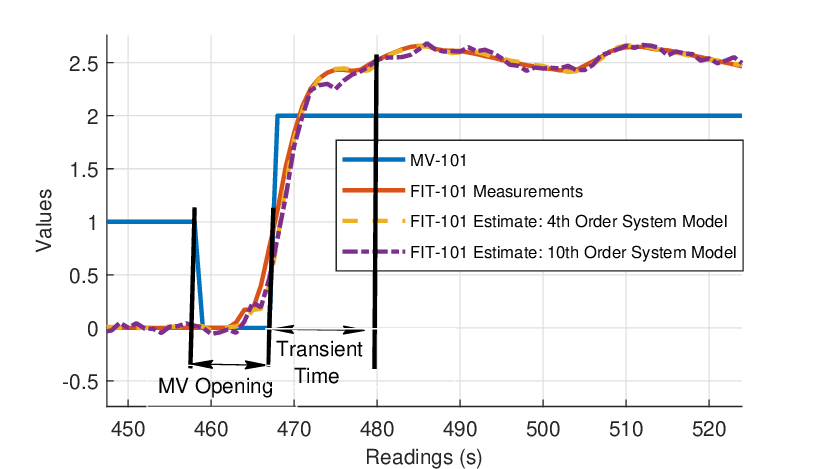}
        \caption{Using two different models obtained via system identification can predict the sensor measurements, capturing the closed loop feedback control and transients, those are robust to small inaccuracies of a system model.}
        \label{fig:sensor_estimation_different_system_model}

        \end{figure}

A common critique of side-channel information-based fingerprinting techniques is that those might fail under a replay attack where an attacker can truthfully reproduce the fingerprint patterns. For an attacker who can learn and spoof the timing profile while spoofing the sensor data, we propose a system model-based watermark technique. Using data collected under normal operation (no attacks) and using the subspace system identification techniques \cite{van1996}, we approximate the input-output dynamical model of actuator-sensor by a set of Linear Time Invariant (LTI) difference equations of the form (\ref{system_model_eq}). Using the system model, we can design an estimator to predict the system state as,



\begin{equation}
\left\{
\begin{array}{ll}
\hat{x}_{k+1} = A\hat{x}_k + Bu_k + L(y_k - C\hat{x}_k), \label{701}   \\
\hat{y}_{k} = C\hat{x}_{k}, \text{ \ \ \ \ \ \  \hspace{6.6125mm} }
\end{array}
\right.
\end{equation}

\noindent where $k \in \mathbb{N}$ is the discrete time index, $\hat{x}_k \in \mathbb{R}^n$ is the state estimate of the approximated model, (its dimension depends on the order of the model), $\hat{y}_k \in \mathbb{R}^m$ are the estimated outputs and $u_k \in \mathbb{R}^p$ denote the control. The system identification problem is to determine the system matrices ${A,B,C}$ from input-output data. We can use the obtained system model to estimate the sensor measurements. This will ultimately help in estimating the \tc by using the Kalman filter. The estimate depends on the internal states in the LTI model, for example, $n=10$ states and $n=4$ states result in different sensor measurement estimates in terms of model accuracy. It is important to validate that the \tc does not deviate too much due to the change in the order of the system model, i.e., it is robust to small model inaccuracies. Figure~\ref{fig:sensor_estimation_different_system_model} shows system model validation for real sensor measurements and estimates under two different models for the flow sensor. It is observed that depending on the order of the model there are slight variations which do not affect the estimates of the \tc. A 4-state system model is provided in the Appendix~\ref{appendix_state_space_model}, for an interested reader. 

\subsubsection{Watermark Design}
The idea of a watermark is to add an external signal to the legitimate input signal so that the added part gets preserved in the output channel, completing a feedback loop. Therefore, the addition of a delay-based watermark in the time at which the command is executed shall be reflected in the sensor measurements.  Designing a practical watermarking approach is considered challenging for an ICS as it can affect the real legitimate input signals. Intuition for our idea comes from the sensor-dependent control and that the process transients are reflected in sensor measurements. A good example is shown in Figure~\ref{fig:dependence_sensor_actuator}.

 
 The key idea is that the control command for opening/closing a motorized valve (MV-101) depends on the level sensor measurements. As soon as the level sensor measures the low water level in the tank, a control command from the PLC is transmitted to open the MV-101. Depending on the time it takes to open the valve, level sensors would show that water is being filled as shown in Figure~\ref{fig:dependence_sensor_actuator}. Our watermark is based on this feedback channel. Therefore, we inject a random delay as a watermark obtained from the PLC clock when the valve opening and closing condition becomes true. This watermark delay shall appear in the \tc measurements done over the sensor data. A replay attacker shall be exposed as it would not know the watermark signal ahead of time. In the case of a replay attack, the device fingerprint obtained from the flow sensor will be preserved, while the watermark signal in the level sensor will expose the attacker. A Watermark signal is generated and recovered inside a PLC and never leaves the system. We used the K-S test to determine if the signal is watermarked or not, that is if it is being replayed.

\begin{figure}
    \centering
    \includegraphics[width=8cm,height=3cm]{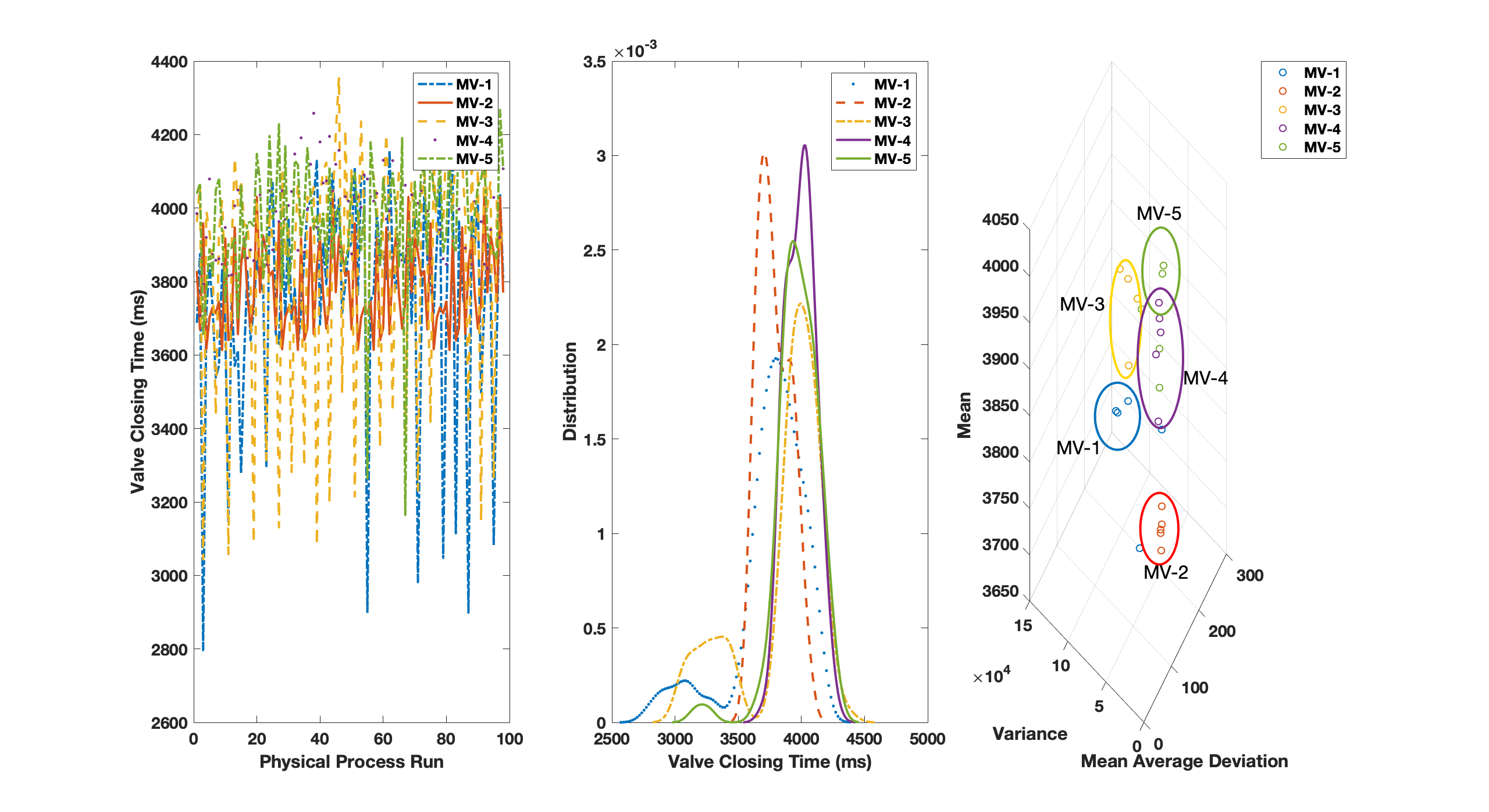}
    \caption{Five identical valves display distinct fingerprints.}
    \label{fig:five_valves_water_loop}
\end{figure}

\subsection{Preliminary Analysis}
Before moving on to the SWaT testbed, a controlled lab setup is created to validate the hypothesis that the time constant does exist. Five motorized valves of the same type and model (CWX15-CR04 U.S. Solid Ball Valve DN15) are used in a water loop process composed of two tanks as shown in Figure~\ref{fig:water_loop_testbed} (Appendix~\ref{appendix_supporting_table}). A key challenge was to obtain the precise timing information, an interrupt-driven sketch for a flow sensor was written to obtain the time in millisecond resolution. Figure~\ref{fig:five_valves_water_loop} shows a visual depiction of data collected for several hundreds of operations of a valve. Given that the same type of devices were used in the same process, we observe from Figure~\ref{fig:five_valves_water_loop} that all five devices could be identified based on \tc fingerprint. There are a few overlaps but given that the process dynamics also contribute to \tc, we expect that unique process dynamics in real-world applications will produce significantly different fingerprints. Next, we will work on a more realistic real-world SWaT plant.

\section{Evaluation}
\label{sec:evaluation}

\subsection{Experimental Setup: SWaT Testbed}

Actuators from a real-world water treatment testbed are used as a case study in this work. Specifically, we have worked on the motorized valves and water pumping motors in a live and realistic functional water treatment plant. Figure~\ref{fig:swat_pic} (Appendix~\ref{appendix_supporting_table}), shows a pictorial view of the SWaT testbed. A total of four devices including two motorized valves (called MV-101 and MV-201) and two pumps (called P-302 and P-602) from the different stages of SWaT are used. The selection of the actuators is made on the criteria of having an associated sensor. We venture to use \tc based fingerprinting to distinguish among all the actuators together and the actuator state for the same actuator. PLCs in the SWaT testbed are used for the design and testing of the proposed watermark-based command validation. 


Support Vector Machine (SVM) is used to train a machine learning model to distinguish between the different actuators based on the \tc fingerprint. LibSVM~\cite{libsvm}, a Python-based library is used. LibSVM~\cite{libsvm} uses the Radial Basis Function (RBF) as a kernel function by default. However, SVM supports more classification functions namely Linear, Polynomial (degree
3), or Sigmoid. If the dataset is linearly separable then the linear model should be enough for analysis. An experiment was performed to show which kernel function should be used to get models with the best accuracy. A chunk size of $10$ readings was selected and a $5$ fold cross-validation was used as a test of classification functions. Our classifier choice is driven by the following reasons: first, the dataset is
{\em small} as only a few samples per actuator/sensor are likely to be
available for training. Second, the dataset is {\em sparse} due to the
dimensionality of the \tc vector. Thus a classifier like
SVM will have no problem identifying and separating hyperplanes that
maximises the {\em margin of separation} between vectors~\cite{schohn2000less_SVM_betterOnLess_ICML}. Using raw sensor values directly is vulnerable to system noise. System noise arises from environmental reasons, for instance, the level of moisture, the pH value of the water, free ions, and the stray currents arising from that all contribute to small differences within the \tc vector. To mitigate system noise, we apply simple statistical functions over raw sensor outputs and compute the \tc fingerprint over said statistics. We formulated the following research questions:
\begin{itemize}
    \item RQ1: Can actuators be fingerprinted using \tc?
\item RQ2: Is it possible to uniquely identify the state of an actuator based on \tc?

\item  RQ3: Can \tc be used for attack detection on sensors and actuators?

\item  RQ4: Can the Watermark technique be used for replay attack detection?
\end{itemize}

\begin{figure}
    \centering
    \includegraphics[scale=0.28]{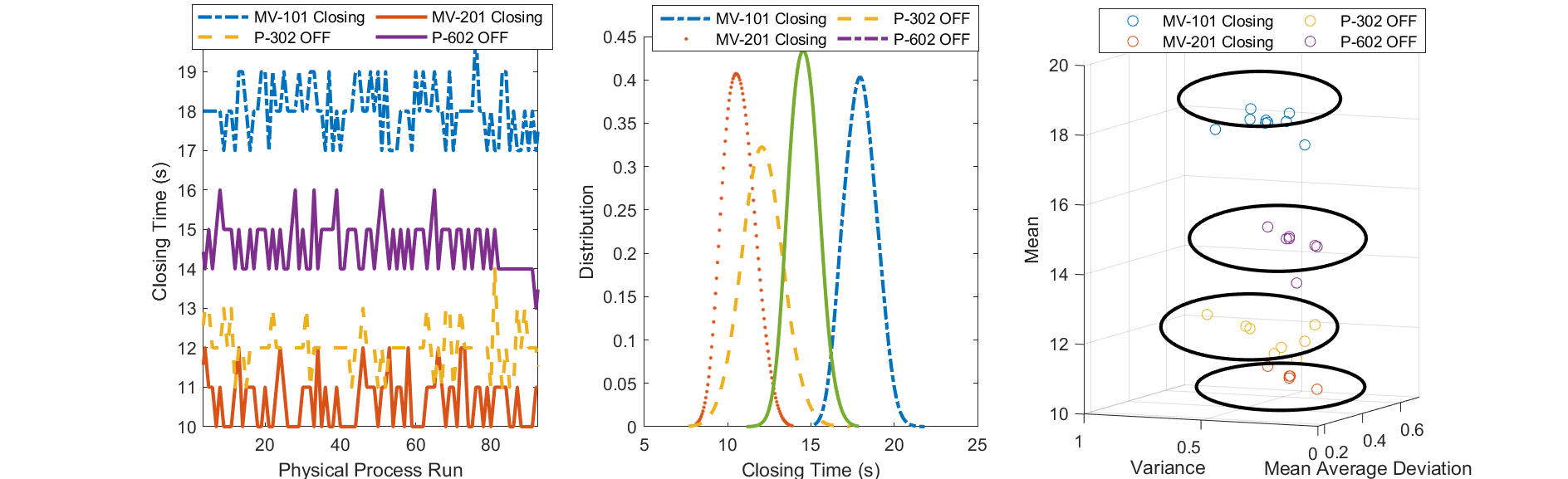}
    \caption{Existence of \tc Fingerprint uniqueness in SWaT. A visual inspection of \tc distribution/clustering of actuators based on three example features. Values of time required for closing valve/pump (left), distribution of closing time (center), and subspace classification of fingerprint vectors (right).} 
    \label{fig:existence_fingerprint}
\end{figure}

\begin{table}
\caption{Performance for different SVM kernel functions}
  \begin{subtable}{.5\linewidth}
    \centering
    \caption{Multiclass classification four actuators.}
    \label{actuator_closing_kernel_function_table}
    \begin{adjustbox}{max width=\textwidth}
 \begin{tabular}[!htb]{|c | c| c | c | c|} 
 \hline
 Kernel Function Type  & Linear & Polynomial & RBF & Sigmoid  \\ 
 \hline
 Accuracy (Closing)  &  100\% &  100\% &  93.87\% &  38.75\%  \\ [1ex]
 \hline
 Accuracy (Opening)  &  100\% &  100\% &  96\% &  38\%  \\ [1ex]
 \hline
\end{tabular}
\end{adjustbox}
  \end{subtable}%
  \begin{subtable}{.5\linewidth}
    \centering
    \caption{Open/ON vs Close/OFF Operation.}
    \label{mv101_openvsclose_kernel_function_table}
\begin{adjustbox}{max width=0.9\textwidth}
 \begin{tabular}[!htb]{|c | c| c | c | c|} 
 \hline
 Kernel Function Type  & Linear & Polynomial & RBF & Sigmoid  \\ 
 \hline
 Accuracy (MV-101)  &  95.65\% &  95.65\% &  91.30\% &  47.82\%  \\ [1ex]
 \hline
  Accuracy (MV-201)  &  100\% &  100\% &  100\% &  50\%  \\ [1ex]
 \hline
  Accuracy (P-302)  &  100\% &  100\% &  97.36\% &  50\%  \\ [1ex]
 \hline
  Accuracy (P-602)  &  100\% &  100\% &  100\% &  50\%  \\ [1ex]
 \hline
\end{tabular}
\end{adjustbox}
  \end{subtable}
\end{table}

\subsection{Actuator Identification (RQ1)}

\noindent \emph{Actuator Fingerprinting: Can actuators be fingerprinted using \tc?}
In Figure~\ref{fig:existence_fingerprint} it can be seen that the total of 4 industrial-grade actuators including two motorized valves and two electric pumps can be uniquely identified using the \tc. In Figure~\ref{fig:existence_fingerprint} we see the \tc values in the left-most graph for the closing/shutting-down process. For different runs of the physical process, \tc values are extracted and displayed on the y-axis. In the middle plot, \tc distributions are plotted for the same closing process of the four actuators. On the right-hand plot, we see that using the three features (only for visual inspection, while SVM used all eight features) on \tc it is possible to cluster each device and hence create a fingerprint. It is to be noted that the motorized valves are of the same type and model and so are the electric pumps. The existence of such a fingerprint shows that it is possible to distinguish between these different devices. 
Next, we endeavor to distinguish among all four devices for the closing/OFF and opening/ON process to see how accurately we can distinguish all of them in a multiclass setting. Table~\ref{actuator_closing_kernel_function_table} shows the result for actuator identification. It turns out Linear and Polynomial kernel functions perform very well since the data is linearly separable. This is in line with the visual result in Figure~\ref{fig:existence_fingerprint}.

\begin{figure}
    \centering
    \begin{subfigure}{0.25\textwidth}
        \centering
        \includegraphics[width=\textwidth]{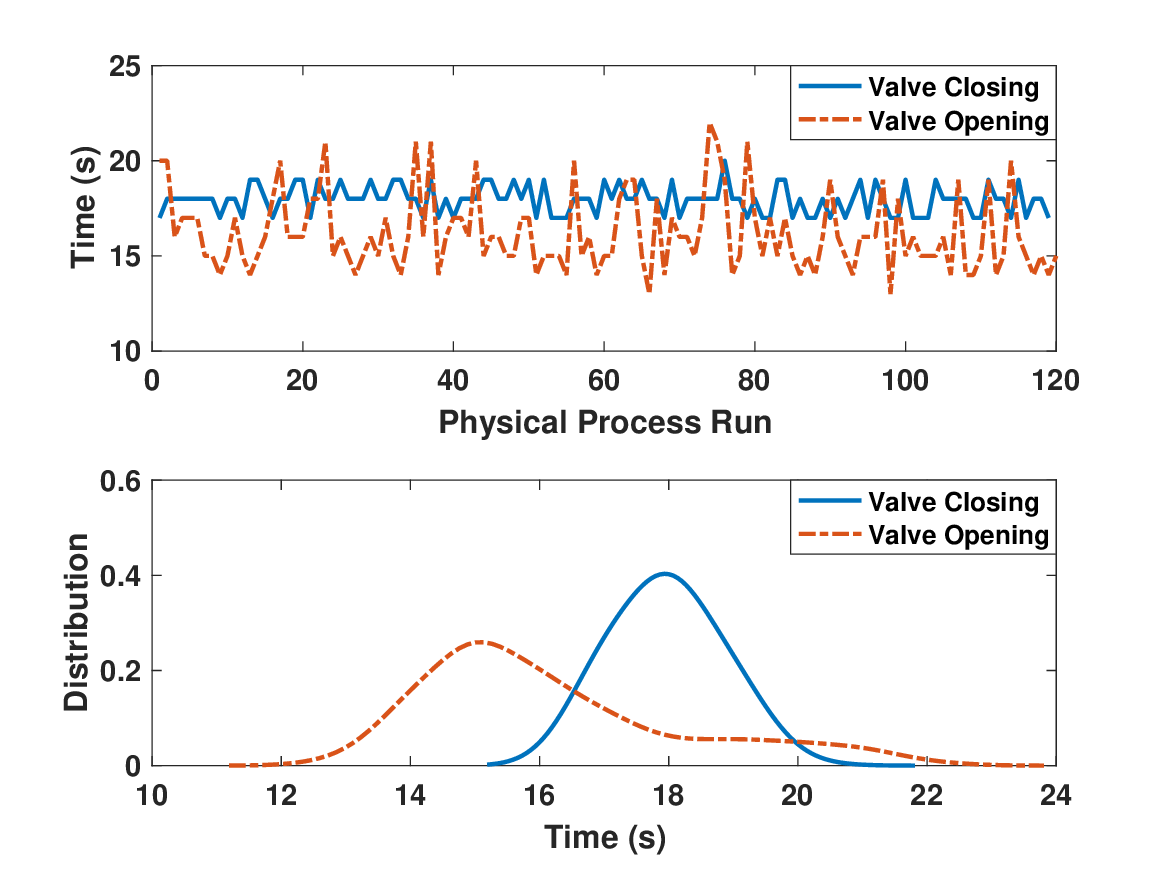}
        \caption{MV opening vs closing.}
          \label{fig:open_vs_close_distribution}
    \end{subfigure}
    \begin{subfigure}{0.25\textwidth}
        \centering
        \includegraphics[width=\textwidth]{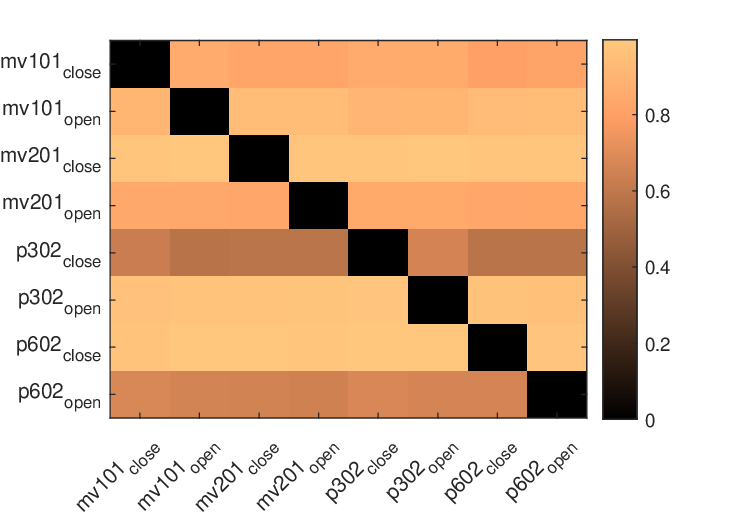}
         \caption{Conditional entropy }
    \label{fig:cond_ent}
    \end{subfigure}
    \caption{(a) timing distributions distinguishes between these two states of the process. The closing process takes more time on average. (b) across individual process fingerprints suggest mutual independence, indicating the uniqueness.}
        \end{figure}



\subsection{Actuator State Fingerprinting (RQ2)}

\noindent \emph{Process State Fingerprinting: Can opening vs closing process of a valve-actuator system be distinguished using \tc?}
Figure~\ref{fig:open_vs_close_distribution} shows the \tc profile distribution for the opening and closing action of a motorized valve. It can be observed that it is possible to distinguish between the two states of the process based on the timing profile. This is an important characteristic of the proposed technique because it is non-trivial to find out what is the real state of the system before, during, or after the attack. An interesting observation is that it takes more time for the valve to close than to open due to different pressure levels across a valve~\cite{ButterFly_Valve_Modelling_2011}, as supported by the formulation in Section~\ref{sec:sources_of_fingerprint}. 
Table~\ref{mv101_openvsclose_kernel_function_table} shows that with high accuracy it is possible to distinguish between the opening and closing operation of the motorized valves and On/OFF operation of the pumps.

\subsubsection{Uniqueness of \tc}

In order to understand the robustness of the extracted feature and understand mutual dependencies, we did an investigation using information-theoretic measures. We study conditional entropy measures across time constants of different processes. It is imperative to study that {\tc} are information-theoretically unique in order to negate the possibility of impersonation attacks. An attacker can use timing information of her processes to design compromises. Let $w(t)$ be the signal corresponding to a \tc. In order to present an information-theoretic analysis on the top, we study the justification of two important criteria- (a) mutual information between features of {\tc} as recorded for the same process, i.e., in successive operations should be high, $\approx$ 1, and (b) conditional entropy of features from a process with that of other process's  {\tc} features should also be $\approx$ 1.

In order to investigate these relations mutual information, $I(\cdot)$, for process $i$ is defined as
\begin{align}
H(w_{ij}|w_{ik})=H(w_{ij})-I(w_{ij},w_{ik}),
\end{align} where $i \in 1:S, j \in 1:N , H(w_{ij})$ is the entropy of $j$th attempt by a process $i$ and $H(w_{ij}|w_{ik})$ is conditional entropy of $i$th process for $j$th attempt, given the features of $k$th attempt. For high recall, mutual information for each of the processes {\tc} should be close to 1 (normalized). Similarly, an ICS process $i$ should not have access to any extra information about process $t$ given observations of its own. Mathematically this can be quantified in conditional entropy as $H(w_{ij}|w_{tk}) \to 1 \ \text{for} \  i,t \in 1:S \ \text{and} \ j,k \in 1:N$. We evaluated entropy measure and mutual information for each of the processes {\tc} as proposed in \cite{conditional}. As can be seen in Figure \ref{fig:cond_ent}, conditional entropies across 8 processes  {\tc} are fairly high, $>0.85$, which supports the use case. The entropy of each of the \tc vectors was recorded to be $\geq 0.9$ . The investigation of conditional entropy across different processes of the ICS system reveals that features are independent.

\subsection{Attack Detection Analysis (RQ3)}

\noindent \emph{How does \tc perform for attack detection on sensors and actuators?} We evaluate our \tc methodology by incorporating it in a CUSUM (cumulative sum)-based anomaly detector. We then test the detector in the live SWaT environment while subjecting the water treatment plant to attacks. 

\subsubsection{Detector Background}
First, the detector tracks, in real-time, actuator ON/OFF operations from the instant of change in actuator state to the time at which the actuator’s associated sensor in each actuator-sensor pair reaches a predetermined threshold that signifies the completion of the ON/OFF operation. 
Sensor thresholds are discussed in subsection~\ref{sec:detector_sensor_value_thresholds}.
Next, the measured transition time is passed to a CUSUM detector. The CUSUM detector tracks changes in transition time of each actuator across consecutive ON/OFF operations, with a separate detector assigned for each combination of actuators and each of the ON/OFF operations, and raises an alarm when an actuator’s transition times for a given operation excessively deviate.


\subsubsection{Sensor Value Thresholds}
\label{sec:detector_sensor_value_thresholds}

For each actuator-sensor pair, we calculate the maximum and minimum associated sensor values ($s_{a, max}$ and $s_{a, min}$ respectively) observed, and then compute the thresholds for the actuator’s ON/OFF operations as,
\begin{equation}
\left\{
\begin{array}{ll}
T_{a, ON} = 0.9 \times s_{a, max} + 0.1 \times s_{a, min}, \label{sensor_threshold_eq} \\
T_{a, OFF} = 0.1 \times s_{a, max} + 0.9 \times s_{a, min}.
\end{array}
\right. 
\end{equation}

\noindent Where $a$ is the actuator under tracking. Transition time is thus defined, for ON operations, as the amount of time between $a$ changing from the OFF state and associated sensor value $y_a \geq T_{a, ON}$ being first true, and for OFF operations as the amount of time between $a$ changing from the ON state and $y_a \leq T_{a, OFF}$ being first true. We illustrate this definition in Figure~\ref{fig:detector_detection_threshold} (Appendix~\ref{appendix_supporting_table}), where the transition time is represented by the horizontal length of the area highlighted in red. We illustrate CUSUM formulation and an example of the CUSUM detector iteration process in Appendix~\ref{appendix:detector_cusum_iteration}.

\subsubsection{Attack Launch and Detection}
\label{sec:detector_experiment_setup}



\begin{figure*}
    \centering
    \begin{adjustbox}{max width=\textwidth}
    \begin{tabular}{ccc}
        \subfloat[]{\includegraphics[width=0.3\textwidth]{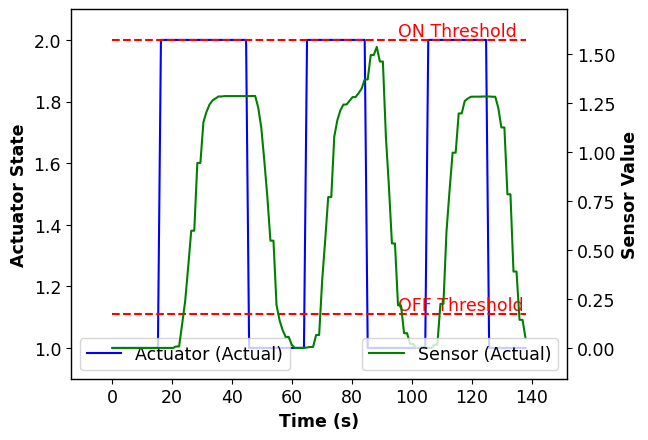}} &
        \subfloat[]{\includegraphics[width=0.3\textwidth]{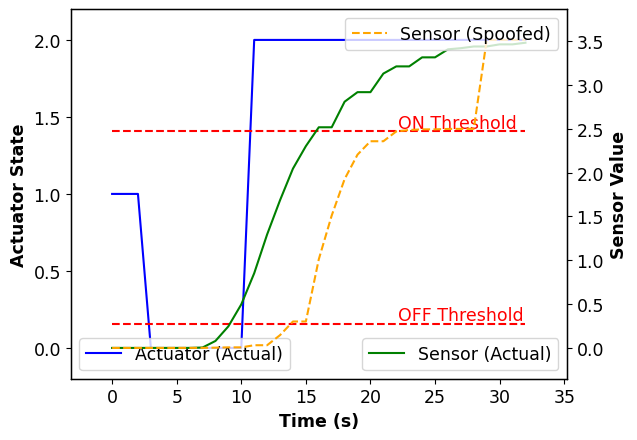}} &
        \subfloat[]{\includegraphics[width=0.3\textwidth]{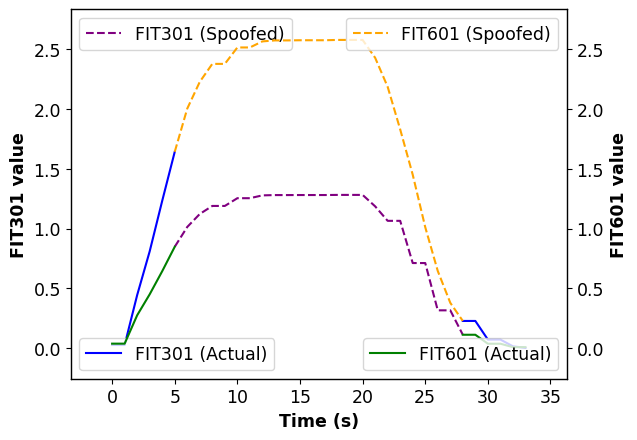}}
    \end{tabular}
    \end{adjustbox}
    \caption{Attacks. Red lines represent ON (top) and OFF (bottom) thresholds as described in subsection~\ref{sec:detector_sensor_value_thresholds}. \emph{(a)} Attack type C1, where an actuator is turned ON and OFF repeatedly. \emph{(b)} Attack type D1, where, during an actuator's ON/OFF operation, the associated sensor is spoofed to follow a sigmoid pattern (in this case between $time = 3$s and $time = 28$s). \emph{(c)} Attack type F1, where the sensors of 2 values are swapped (in this case between $time = 5$s and $time = 28$s).}
    \label{fig:detector_atk_type}
\end{figure*}

In the online phase, we set up the detector to receive data on the status of actuators and sensors in SWaT from SWaT’s historian at a frequency of 1 Hz. Next, we ran SWaT with the detector connected for a total of 10 hours over multiple sessions, while simultaneously running a script that launched brief attacks on the plant at a maximum frequency of 1 Hz. 
A list of all the types of attacks performed on the plant can be found in Table~\ref{table:detector_attack_type_table}, and a few attack types are shown in Figure~\ref{fig:detector_atk_type}. 

\begin{table*}[!ht]
\caption{List of attack types used in evaluating CUSUM-based anomaly detector}
\label{table:detector_attack_type_table}
\centering
\begin{adjustbox}{max width=\textwidth}
\begin{tabular}{|l|p{\textwidth}|}
    \hline
    Type & Description \\
    \hline
    A1 & One or more flow sensors report spoofed constant values simultaneously \\
    \hline
    B1 & One or more actuators are set to manual mode, and one command is sent to each actuator, for a pre-configured period \\
    \hline
    C1 & The attack script toggles one or more actuators on and off repeatedly for a given period \\
    \hline
    D1 & The attack script waits for (or manually triggers) an actuator to perform an ON/OFF operation, for which the corresponding sensor is spoofed during the operation’s transition time to follow a sigmoid pattern \\
    \hline
    D2 & The attack script waits for (or manually triggers) an actuator to perform an ON/OFF operation, for which the corresponding sensor is spoofed during the operation’s transition time to constantly report its initial value at the start of the operation \\
    \hline
    E1 & The status of an actuator and sensor readings for an actuator-sensor pair are spoofed simultaneously to mimic an ON/OFF operation of the actuator with the sensor values following a sigmoid pattern \\
    \hline
    F1 & The readings of two sensors are swapped for a given period \\
    \hline
\end{tabular}
\end{adjustbox}
\end{table*}

\begin{table}[!htp]
\caption{Attack detection for each type of attack.}
\label{table:detector_attack_detection_rate}
\centering
\begin{adjustbox}{max width=0.48 \textwidth}
\begin{tabular}{|p{0.24\textwidth}|c|c|c|c|c|c|c|}
    \hline
    \raggedright Detection Stats / Attack Type & A1 & B1 & C1 & D1 & D2 & E1 & F1 \\
    \hline
    \raggedright Number of Attacks Performed & 29 & 116 & 105 & 53 & 62 & 117 & 39 \\
    \hline
    \raggedright Overall Detection Rate (\%) & 17.24 & 52.59 & 100.00 & 83.02 & 96.77 & 88.89 & 30.77 \\
    \hline
    \raggedright\textbf{CUSUM Detection Rate (\%)} & \textbf{13.79} & \textbf{51.72} & \textbf{96.19} & \textbf{83.02} & \textbf{95.16} & \textbf{54.70} & \textbf{28.21} \\
    \hline
    \raggedright Detection Rate-Incomplete Operations (\%) & 13.79 & 29.31 & 90.48 & 0.00 & 74.19 & 34.19 & 17.95 \\
    \hline
    \raggedright Detection Rate-Timed Out Operations (\%) & 0.00 & 0.86 & 0.95 & 0.00 & 1.61 & 28.21 & 0.00 \\
    \hline
\end{tabular}
\end{adjustbox}
\end{table}

Table~\ref{table:detector_attack_detection_rate} illustrates our anomaly detector's rate of attack detection for different attack types. The results demonstrate that our CUSUM anomaly detector shown in bold in Table~\ref{table:detector_attack_detection_rate} is effective in detecting attack types C1, D1, D2 and E1 with an overall detection rate of $\geq 80\%$ for each type, and less effective in detecting other types. Generally, CUSUM-based detection facilitated the greatest detection rate, while detection through timed-out operations facilitated the lowest detection rate, for all attack types.
Attack types A1 and F1 do not involve ON/OFF operations of actuators, whether spoofed or actual. As such, without such operations, it was expected that our detector would fail to detect instances of these attack types as there is no transition period to observe. These expectations are reflected in our results. Conversely, attack types B1, C1, D1, D2 and E1 all involve ON/OFF operations and thus we expected to see high detection rates for these types of attacks, except for B1 for which we expected a lower detection rate. 

For attack type C1, as seen in Figure~\ref{fig:detector_atk_type} \emph{(a)}, given a sufficiently high frequency of toggling of an actuator under attack, the associated sensor would not be able to reach the sensor value thresholds for certain ON/OFF operations before the actuator is toggled again. This is seen in Figure~\ref{fig:detector_atk_type} \emph{(a)}, where, after pump P-602 was first turned on at about $time = 15$s, FIT601’s value failed to reach the ON operation threshold before pump P602 was turned off for the first time at about $time = 45$s. This may cause the transition period for other ON/OFF operations to be abnormally short, as was the case with P602’s OFF operations in the figure, which our detector was expected to raise an alert for.

For attack types D1, D2, and E1, we assumed that a spoofing attacker has knowledge of neither the sensor thresholds used for measuring transition periods of ON/OFF operations nor the CUSUM parameters used for evaluating transition periods (including means of transition periods), and so we expected our detector to successfully detect spoofing attacks where the attacker has spoofed the sensor to follow patterns reflecting transition periods that are too short or too long.
For attack type B1, while ON/OFF operations are also involved in attacks of this type, no spoofing of sensors is involved. 
One post-experiment observation made was that attacks of this type were carried out even if the actuator command sent as part of the attack would not cause any ON/OFF operation to occur due to the targeted actuator already being in the desired state. Regardless, the observed detection rate is higher than expected. In the next section, we consider even stronger attackers that have \tc knowledge and try to launch replay attacks.

\begin{figure}
    \centering
        \includegraphics[width=0.35\textwidth]{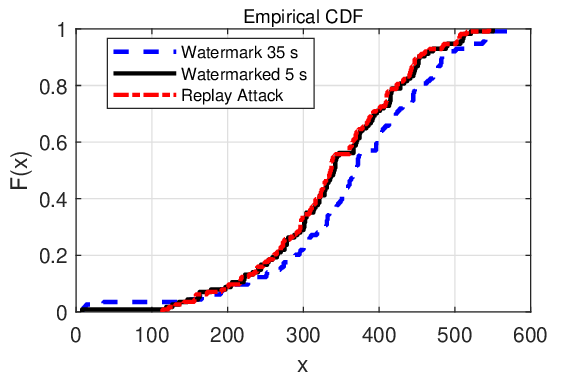}
    \caption{Empirical CDF: Normal data being replayed and two watermark signals. It is empirically found to successfully detect a replay attack using the K-S test, for a watermarked delay of more than 35 seconds.}
    
       \label{fig:watermark_lit101}
        \end{figure}

\subsection{Watermark Assessment (RQ4)}

\noindent \textbf{Definition} Let sensor measurements under a replay attack be  $y_k^a$, control command under replay attack  $u_k^a$, and the state estimate  $x_k^a$ at  time step $k$,  $0 < k \leq T$ for an attack time period $T$.

\noindent \textbf{Definition} {\em  Watermark $\Delta k$}:  The output, i.e. the transient response in sensor measurement, depends on the control command execution time, triggering the actuation action. A watermark is added to the control input such that the effects of the added watermark are observable on the output of the system $y_k$. 
\noindent \textbf{Theorem 4.1} Given the system model of \eqref{701}, a relationship for input watermarked control and sensor measurement can be derived as, $    y_{k+1} = CAx_{k} + Cv_{k} + \eta_{k+1} + CB (u_{k - \Delta k})$.

\noindent \emph{Proof:} Proofs are provided in the Appendix~\ref{appendix_theorem_proof}.







\noindent A detailed analysis of quantifying the watermark signal is done using the K-S test. A K-S test is used to distinguish between the base distribution and a watermarked distribution. The K-S statistics quantifies the distance between the empirical Cumulative Distribution Functions (CDF) of two samples. 
\begin{equation}
        D_{n,m} =  \sup_{x \in \mathbb{R}} | F_n(x) - G_m(x) |.
\end{equation}

\noindent The results from our experiment using watermark signal on the stage~1 of the SWaT testbed are shown in Figure~\ref{fig:watermark_lit101}. Figure~\ref{fig:watermark_lit101} shows the empirical CDF for \tc of the motorized valve MV-101. It shows distributions for a replay attack, and two different watermark delays (5 and 35 seconds). $D_{n,m}$ is the distance between any of the two distributions as explained above. From Figure~\ref{fig:watermark_lit101} it can be seen that it is easier to distinguish between replayed data and a watermark delay of 35s however, the 5s watermark signal is hard to distinguish visually due to the process and measurement noise components. In the following, we precisely model the watermark component.

\begin{table}[!htp]
\caption{Critical state modes of operation for the two stages in SWaT testbed. T:TRUE and F:False, means the physical states are close to either critical state or not, respectively.}
    \centering
  \label{table:critical_states}
    \scalebox{0.7}{
\begin{tabular}{|c|c|c|c|c|c| }
\hline
   Mode & $X1_{iH}$ & $X1_{iL}$ & $X2_{iH}$ & $X2_{iL}$ & $T_q$  \\
    \hline
  1 &     F & F &F &T & $T2_{qL} = \frac{\Delta_{L2}}{R2_{out}}$ \\ 
    \hline
    2 &    F & F & T &F & $T2_{qH} = \frac{\Delta_{H2}}{R2_{in}}$ \\ 
    \hline
3 &        F & T &F &F & $T1_{qL} = \frac{\Delta_{L1}}{R1_{out}}$ \\ 
    \hline
  4 &   F & T & F & T & $min(T1_{qL},T2_{qL}) \leq T_q \leq T1_{qL} + T2_{qL}$ \\ 
    \hline
5 &     F & T & T & F &  $min(T1_{qL},T2_{qH}) \leq T_q \leq T1_{qL} + T2_{qH}$ \\ 
    \hline
  6 &   T & F & F & F & $T1_{qH} = \frac{\Delta_{H1}}{R1_{in}}$ \\ 
    \hline
7 &     T & F &F & T & $T_q = min (T1_{qH}, T2_{qL)}$ \\ 
    \hline
 8 &    T & F & T & F & $min(T1_{qH},T2_{qH}) \leq T_q \leq T1_{qH} + T2_{qH}$  \\ 
    \hline
\end{tabular}}
\end{table}

\subsubsection{Bounds on Watermark Delay}
Time to critical state~($T_q$) has been used as a metric for defensive solutions in an ICS ~\cite{urbina_CCS2016limiting,Carlos2017reachable_TimeToCriticalState,John2019modular_ACNS_TimeToCriticalState}. A critical state is defined as the state of the physical process that is out of safety bounds, causing harm or degrading the system performance. This metric, being a measure of time,  is a good fit for choosing the watermark to be added, while keeping the system performance and safety within the defined bounds. Let us define the system's initial state as $X_i$, the critical state as $X_q$, the maximum rate at which system is moving towards the critical state as $R_q$, and distance between the initial state to the critical state as $\Delta_q = X_q - X_i$ then $T_q$ can be measured as,
\begin{equation}
    T_q = \frac{\Delta_q}{R_q}, \forall {q \in Q(critical-states)}.
\end{equation}

\noindent \emph{Example:}  To understand the idea, consider the example shown in Figure~\ref{fig:stage1_diagram}, e.g., one such critical state can be to overflow the tank. We can measure the time required from a current safe state to reach a critical state and in turn, use that time as bound to add the watermark. From the design of the water treatment plant~\cite{swat2016}, it is known that each process has defined limits. For example, water tanks have high (H) and low (L) level limits, that should not be crossed to keep the system in a safe operating range. In this case for LIT-101, $H=800 mm$ and $L=500 mm$; by design, these are the levels at which the actuators commands are triggered -- authentication of which is the subject of this study. Earlier studies~\cite{John2019modular_ACNS_TimeToCriticalState,Eyasu2020cima_timeToCriticalState} need to estimate the time to critical state based on the initial system state, hence, leading to uncertain time measurements. However, by definition, we only need to consider the worst case, i.e., the water level is $H$ and water ingress did not stop, making $H$ a best starting point for an attack, therefore, a worst-case for defense. The critical state for Tank-1 in Figure~\ref{fig:stage1_diagram} is defined as $1000 mm$. Therefore, $\Delta_{H1} = 1000 - 800$ is the buffer to handle the impact before reaching the critical state, i.e., tank overflow. Imagine that the LIT-101 level reading is $800mm$ and a decision is taken not to turn OFF MV-101 valve at the inlet, then depending on the rate-in $R1_{in}$ and rate-out $R1_{out}$ of water flow we can calculate the time to overflow as $T1_{qH}$,         
\vspace{-3mm}
\begin{equation}
    T1_{qH} = \frac{\Delta_{H1}}{R1_{in} - R1_{out}},  R_q =\{R1_{in}=max, R1_{out}=0\}
\end{equation}\vspace{-3mm}

Similarly, $T1_{qL}$ is the time required to underflow the tank. For the single tank example, it is straight forward calculation given the system design parameters. However, it is not hard to establish the measurement for the multiple stages of a physical process. For the water treatment system, the next stage's initial states are labeled as $X2_{iH}$ and $ X2_{iL}$, for high and low levels, respectively. Table~\ref{table:critical_states} shows all the (physically) possible initial states for a two-stage tank system. Consider the last row in the table where $T2_{qH}$ is limited by $\Delta_{H2}$, whereas, $T1_{qH}$ get even more buffer time as the tank-2 is being filled from tank-1, meaning it shall take more time to overflow tank-1 when $R1_{out \neq 0}$. However, we are considering the buffer zone for each stage only in our design, considering a worst-case. From the plant design and dataset, the rates are calculated~\cite{John2019modular_ACNS_TimeToCriticalState} as, $R1_{in}=0.48mm/s$ and $R1_{out}=0.47mm/s$. This results in $T1_{qH} = 416.66 s$ in the worst-case scenario and $T1_{qL} = 744.68 s$ for underflow, given $X1_{iL}=500 mm$ and $X1_{qL}=150 mm$. This provides us with a $T1_q =\{416.66,744.68\}$(s), which is a reasonable range to chose a watermark delay from, without bringing the system into a critical state.  

We can expand the idea of the watermark to multiple stages of a physical process to expand the state-space to chose a delay. Table~\ref{table:critical_states} shows the physically possible combinations of critical states in the two stages of SWaT presented as different operating modes. For example, consider the modes $4,5$ and $8$, it is possible to extend the time to a critical state by considering the cascading effect of the delay. On this end, note that as a defender it is possible to design a validation phase where a desired mode of operation is activated involving the multiple stages of a physical process. This approach increases the delay-space to chose the nonce from.

\begin{table}[!t]
\centering
   \caption{NIST Statistical Test Suite Results shows sufficient randomness in the generated watermark.}
    \label{table:nist}
\scalebox{0.7}{
\begin{tabular}{|l|l|}
\hline
\textbf{Test} & \textbf{p Value} \\ \hline
Runs & 0.92 \\ \hline

Longest run of 1's     & 0.82               \\ \hline
Block of Freq.   & 0.64 \\ \hline
Runs  & 0.78         \\ \hline
Cum. sums (Fwd, Rev)  & 0.48, 0.53       \\ \hline
Approx. Entropy & 0.87       \\ \hline
Serial & 0.54        \\ \hline
Frequency  & 0.63       \\ \hline
\end{tabular}}
\end{table}

\subsubsection{Randomness Analysis on Watermark}

 Further, NIST statistical test \cite{nist} was conducted to study randomness in the watermarks and the p-values are shown in Table \ref{table:nist}.  The p-value
corresponds to the probability that a perfect random number generator would have produced a sequence less
random than the sequence put to test. A p-value larger than 0.01 suggests that the generated bits are random with a confidence of 99\%. The analysis and p-values suggest that there is sufficient randomness in the generated secret from the watermarks. 
 


\subsubsection{Quantifying the Watermark}

While modeling the system in a state-space form, the system states can be estimated using the Kalman filter. This formulation helps in detecting the delay added in the control input that is reflected in the sensor output and is useful in quantifying the contribution of the watermark signal in the sensor measurements by normalizing the input and output in terms of the residual. The residual is defined as the difference between the sensor measurements and sensor estimation.


\noindent \textbf{Theorem 4.2} Given the system model   in~\ref{701}, Kalman filter gain $L$, and watermarked signal $\Delta k$, it can be shown that the difference between output signals is driven by the watermark signal and can be given as, $r_{k+1} =(CA - CLC)(x_k^a - \hat{x}_k^{wm})+ \eta_{k+1} + Cv_k    + CB(u_k^a - u_{k-\Delta k})$.

\noindent \emph{Proof:} Proof is provided in the Appendix~\ref{appendix_theorem_proof}. The term $CB(u_k^a - u_{k-\Delta k})$  quantifies the effects of watermark. An attacker's goal is to issue a control command $u_k^a = u_{k-\Delta k}$, to make this term zero and drive $E[r_k]\xrightarrow{}0$, as if it's in the normal operation.

\subsubsection{Security Argument}

To understand this argument, it is critical to consider how does watermark affects the physical process. Without any watermark, an adversary could record the process data beforehand and then replay that to a SCADA system. The adversary was able to do that because of the repetitive and periodic nature of the process data. However, with the addition of a watermark, now an adversary can not just record the data a priori and then start sending that to the system operator, hiding the real physical system state. For example, continuous pumping would result in water level dropping to a defined $low$ set point, at which the pump shall stop, however, we chose (under watermarking) to stop the pump at random and not at a predefined point. An attacker on the network needs to precisely find this random stopping time and also predict the sensor measurements to stay stealthy. To do this, assume the attacker relies on a change-point detection using the observations on $L$ separate sensors and actuators. Attacker records a signal $X_{l,n}$ at time $n$ at the device $D_l \in L$. A change occur at an unknown point in time $\lambda$, $\lambda \in \{1,2,...\}$, rendering $X_{l,n}$ probability distribution as $p_0$($x$) for $n < \lambda$ and $p_1$($x$) for $n \geq \lambda$.

\noindent \textbf{ Theorem 4.3} For an attacker trying to detect changes in the physical observations due to watermark, the detector shall have a false alarm probability equal to $\alpha$. Due to noise in the observations, there is a non-zero probability for the attacker's detection delay $\tau > 0$. That the attacker will be able to see the change and spoof the signal after $\delta$ time difference, from the start of change at $\lambda$ time. 
\noindent \emph{Proof}: The proof is based on the results reported on point-change detection~\cite{blazek2001novel_PointChangeDetection,tartakovsky2005general_pointChangeDetection}, especially in the context of random physical challenge~\cite{shoukry2015}.  By challenging a sensor at a randomly selected time $\lambda$ and an attacker who is aware of such a challenge, but at the same time is spoofing the true measurements, therefore needs to consistently reflect the anomalous profile starting at time $\lambda$. The attacker's detector needs to wait $\delta$ seconds to recognize that the probability distribution has changed to $p_1(x)$. This $\delta$ delay in the attacker's response can be leveraged to detect incoherent responses to the watermark.  \hfill $\blacksquare$

\section{Discussion}
\label{sec:discussion}




By the results seen so far it can be concluded that the proposed watermarking on top of \tc can deter a replay attack. In a standard replay attack, an adversary shall record past sensor-actuator measurements to replay but added watermark would not reflect in the replayed sensor measurements, tipping off the presence of an adversary. Note that the replay attack represents a worst-case scenario and it is not the only type of attack the proposed technique can detect. For example, \tc would be an ideal solution for attacks specifically taking advantage of transient states, as in the case of power generators as discussed in Appendix~\ref{sec_use_case_EPIC}. The idea of this work relies on precise mathematical models of the physical process, aided with the idea of physical challenge in the form of a watermark and physical properties namely \tc. Randomness in the watermark is a key aspect, considering a single-stage process, it has been shown that in a worst-case scenario we have more than $400 s$ to choose the watermark delay. It was also shown that by increasing the number of stages, the delay-space also becomes bigger and the complex interactions among different stages make it even harder for an attacker to correctly predict the randomness and do it right for all the stages involved at the same time. Precisely, important parameters are the sampling rate at which the defender can receive the data and the granularity of the randomness, e.g., a resolution of micro-seconds would provide a much bigger state-space as compared to minutes scale. As a defense, both these parameters can be made a part of the design of the process. Moreover, SVM as a classifier works well with fewer samples. Unlike SVMs, DNNs require large training datasets whereas we must characterise the \tc with just a few samples in an ICS environment. DNNs are appropriate for high-dimensionality settings such as image classification where the vector sizes is tens of thousands as opposed to \tc with a vector size of ~100. Instead of using 100 dimensional vectors, we use eight features which further reduces dimensionality.

\noindent \emph{Ageing Effects}: An important aspect of a device fingerprinting technique is, how stable it is with an ageing device. This is intuitive that any fingerprint pattern that depends on hardware would change over time. \tc also depends on the hardware features but an interesting aspect of the proposed technique is the watermark signal that is designed based on the timing channel. The watermark signal shall be free of these ageing effects as the watermark is generated in the control logic and shall be able to compensate for device ageing effects.


\noindent \emph{Real-world Implementations}:
We have demonstrated the effectiveness of the proposed technique on a real water treatment system. Watermark-based nonce never leaves the system. It is generated in the PLC and recovered from the sensor data inside the PLC. This is used for proof of data liveness. It uses one system call in ladder logic of the PLC to add the watermark and then for recovering response it takes a comparator instruction measuring the number of scan cycles it took until reaching the particular thresholds to calculate \tc. A system model helps in estimating the watermark effect in sensor measurements. We can obtain simple models for each sensor-actuator pair and that reduces the matrix multiplication to a scalar multiplication for state estimation, which can be performed inside a PLC using a structured text language script.

\section{Conclusions}
We conclude that the actuator and process transient-based feature called as \tc, fingerprints the actuators and the state of an actuator. \tc can be extracted indirectly from the sensor measurement, which proved critical to validate the authenticity of an actuator without actually relying/trusting on SCADA control channels. Earlier works highly depend on the SCADA control channel to extract actuator fingerprint. A watermarking approach designed on top of \tc can tackle the fingerprint preserving attacks, e.g., a replay attack. Indeed, the {\em Watermark-}\tc is a step towards solving the long-standing problem of freshness where authentication is based on the physical properties.




\newpage




\bibliographystyle{ACM-Reference-Format}
\bibliography{AsiaCCS2024_main.bib}

\appendix


\section{CUSUM Description}
\label{appendix:detector_cusum_iteration}

\subsection{Cumulative Sum (CUSUM) Based Detection}
\label{sec:detector_cusum_detection}

Our operation transition times are passed as input to a cumulative sum (CUSUM) procedure which we use as a stateful detector, with a unique instance of the procedure used for each combination of actuator and type of operation (ON/OFF). This use of CUSUM is a means of detecting changes in statistical properties of transition times measured. Each transition time is fed to the corresponding detector as a single scalar as soon as it is measured.


We first define symbols used in explaining the procedure. a: {Actuator under tracking},
$p \in \{ON, OFF\}$: {Type of actuator operation},
$i \in \mathbb{N}_{\geq 0}$: {Time step (with $1^{st}$ iteration at $i=1$)},
$k \in {+, -}$: {Direction of tracked changes},
$S_{a, p, i}^k \in \mathbb{R}$: {CUSUM},
$T_{a, p}^k \in \mathbb{R}$: {CUSUM threshold},
$\mu_{a, p} \in \mathbb{R}^+$: {Mean transition time},
$\beta_{a, p} \in \mathbb{R}^+$: {Bias},
$t_{a, p, i} \in \mathbb{R}^+$: {Measured transition time.}

\noindent Our CUSUM detector~\cite{process_skew} iterates using the following rule:
\begin{equation}
\left\{
\begin{array}{ll}
d_{a, p, i}^+ = S_{a, p, i - 1}^+ + t_{a, p, i} - \mu_{a, p} - \beta_{a, p},\\
d_{a, p, i}^- = S_{a, p, i - 1}^- + t_{a, p, i} - \mu_{a, p} + \beta_{a, p},\\
S_{a, p, i}^+ =
\begin{cases}
\max(0, d_{a, p, i}^+),& \text{if } d_{a, p, i}^+ \leq T_{a, p}^+ \\
0 & \text{otherwise}
\end{cases},\\
S_{a, p, i}^- =
\begin{cases}
\min(0, d_{a, p, i}^-),& \text{if } d_{a, p, i}^- \geq T_{a, p}^- \\
0 & \text{otherwise}
\end{cases},\\
\end{array}
\right.
\label{eq:detector_cusum_iteration_rule}
\end{equation}

\noindent When $d_{a, p, i}^+ > T_{a, p}^+$ and/or $d_{a, p, i}^- < T_{a, p}^-$, the CUSUM detector raises an alarm. 
The mean transition time $\mu_{a, p}$ is derived by taking the average transition time found in a given SWaT dataset for each actuator-operation combination. The bias $\beta_{a, p}$ is derived by halving the standard deviation of each such combination's transition times, and facilitates gradual drifting of the CUSUM values to 0 under normal conditions. The threshold $T_{a, p}^k$ are derived using binary search such that a fixed maximum false alarm rate (in this case 2\%) is reached on the dataset in each direction. 

\subsection{Iteration Example}
In this section we provide an example of the results of performing CUSUM detection on ON transition times of the motorized valve MV101. Table~\ref{table:detector_cusum_iteration_example_params} lists the relevant parameters used for this procedure.

\begin{figure*}
    \centering
    \begin{adjustbox}{max width=\textwidth}
        \begin{tabular}{ccc}
            \subfloat[]{\includegraphics[width=0.33\textwidth]{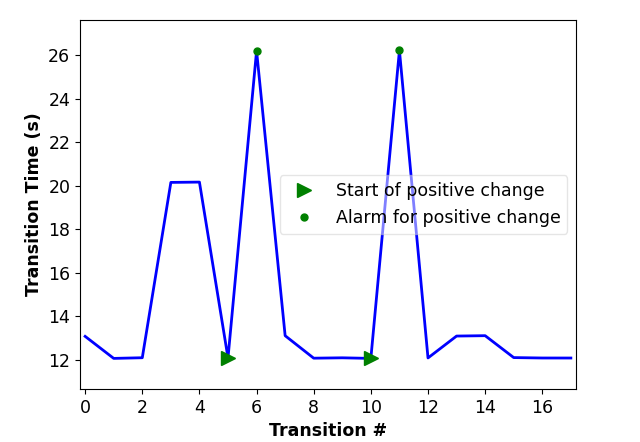}} &
            \subfloat[]{\includegraphics[width=0.33\textwidth]{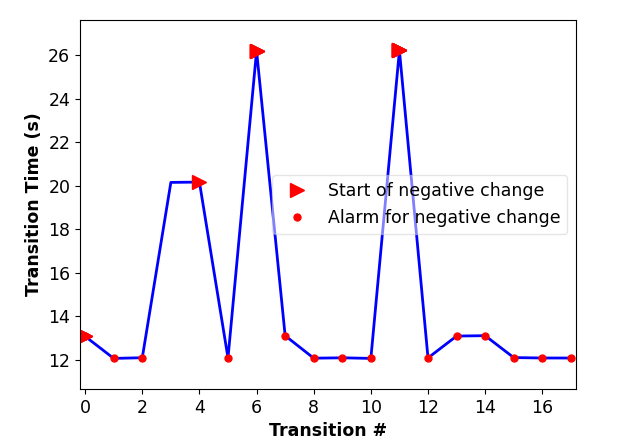}} &
            \subfloat[]{\includegraphics[width=0.33\textwidth]{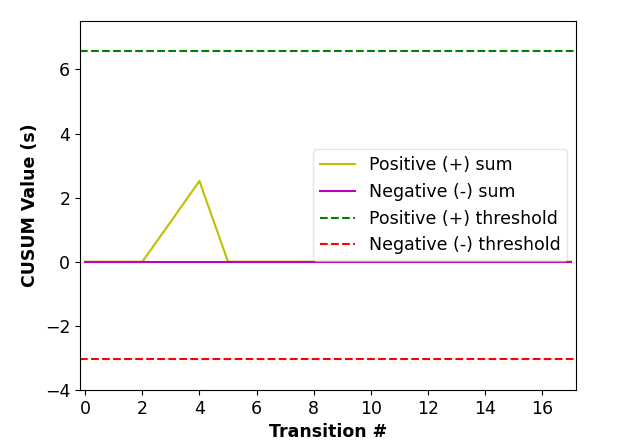}}
        \end{tabular}
    \end{adjustbox}
    \caption{Sample CUSUM iteration results for MV101 ON transition times. \emph{(a)} ON transition times for MV101 along with raised CUSUM alarms for positive CUSUM. \emph{(b)} ON transition times for MV101 along with raised CUSUM alarms for negative CUSUM. \emph{(c)} Positive and negative CUSUM values for each transition time recorded/iteration of the CUSUM procedure.}
    \label{fig:detector_cusum_iteration_example}
\end{figure*}

Figure~\ref{fig:detector_cusum_iteration_example} \emph{(a)}-\emph{(b)} indicate the transition times recorded for MV101's ON operations, as well as the alarms raised for each direction of change. There are 2 alarms raised for positive CUSUM, and 14 alarms for negative CUSUM (with the first negative CUSUM alarm being raised for the first transition time recorded, and the indicator in the graph for said alarm being obscured). We also illustrate the start of alarm-inducing changes. For each such change, the start is defined as the latest iteration preceding the alarm, where CUSUM was set to 0 as a result of the $max$ function for positive CUSUM and the $min$ function for negative CUSUM in Equation~\ref{eq:detector_cusum_iteration_rule}, and the start signifies the likely point at which alarm-inducing deviations in transition time started.

Figure~\ref{fig:detector_cusum_iteration_example} \emph{(c)} illustrates CUSUM values throughout each iteration. It is useful to note that, in Equation~\ref{eq:detector_cusum_iteration_rule}, negative values are rounded to 0 when calculating positive direction CUSUM for each iteration, and that CUSUM values are set to 0 in the case of an alarm being raised, both of which account for the zero CUSUM values found throughout most iterations.

\begin{table}[htp]
    \caption{Key parameters used for our CUSUM iteration example. Refer to Definition~\ref{eq:detector_cusum_iteration_rule}  for definitions of parameters.}
    \label{table:detector_cusum_iteration_example_params}
    \centering
    \begin{adjustbox}{max width=0.4\textwidth}
    \begin{tabular}{|c|c|}
        \hline
        Parameter & Value \\
        \hline
        $\mu_{a, p}$ & 17.79 \\
        \hline
        $\beta_{a, p}$ & 1.12 \\
        \hline
        $T_{a, p}^+$ & 6.56 \\
        \hline
        $T_{a, p}^-$ & -3.05 \\
        \hline
    \end{tabular}
    \end{adjustbox}
\end{table}

\section{State Space System Model}
\label{appendix_state_space_model}
In the following state space matrices, Kalman Filter gain $L$, input/output vectors are shown along with the estimation equation for the case of $n=4$ states. This system model captures the process dynamics of Stage~1 of the SWaT process which is shown in Figure~\ref{fig:stage1_diagram}. Using the following LTI equation~\eqref{701}, given the state space matrices and live data from the plant, one can estimate the sensor measurements as shown in Figure~\ref{fig:sensor_estimation_different_system_model}. 

\noindent\rule{\hsize}{1pt}\vspace{.5mm}
$A=\begin{bmatrix}
    1.0000  &  0.0008 &   -0.0003  &  0.0031 \\
   -0.0026  &  0.9782  &  0.1173  & -0.0037 \\
   -0.0057  & -0.0614  &  0.7645  &  0.3523 \\
   -0.0091  &  0.0030  & -0.0417 &    0.8197 \end{bmatrix}$, 
   
   $B=\begin{bmatrix}
       0.0000  &  0.0000 &  -0.0000 \\
   -0.0003  &  0.0001  &  0.0000 \\
    0.0009  & -0.0007  &  0.0001 \\
   -0.0010  &  0.0004   & 0.0002 \end{bmatrix}$, \\
   
   $u_k=\begin{bmatrix}
     MV-101 \\ P-101 \\ P-102 \end{bmatrix}$, $y_k=\begin{bmatrix}
      FIT-101 \\
   LIT-101 \end{bmatrix}$
   
   $C=1.0e+04 *\begin{bmatrix}

    0.0018  & -0.0128 &  -0.0006 &  -0.0001 \\
   -2.9695  & -0.0029  &  0.0002  & -0.0028
\end{bmatrix}$

$L=\begin{bmatrix}
      -0.0001 &  -0.0000 \\
   -0.0073  & -0.0001 \\
   -0.0282   & 0.0010 \\
   -0.0038  & -0.0020 \end{bmatrix}$



\noindent\rule{\hsize}{1pt}\\[.5mm]

\section{Theorem Proofs}
\label{appendix_theorem_proof}




\textbf{Theorem 4.1} \emph{ Given the system model of \eqref{701}, a relationship for input watermarked control and sensor measurement can be derived as, $    y_{k+1} = CAx_{k} + Cv_{k} + \eta_{k+1} + CB (u_{k - \Delta k})$.}

\noindent \emph{Proof:} The proposed watermarking technique injects a delay defined as $\Delta k$ in the control signal $u_k$. A replay attack will use the normal data and system model as defined in Eqn.\, \eqref{701}. The watermarked input would result in,

\begin{equation}\label{system_state_withwatermark_eq}
    x_{k+1} = Ax_k + B(u_{k - \Delta k}) + v_k, 
\end{equation}

\begin{equation}\label{output_withwatermark_eq}
    y_{k+1} = Cx_{k+1} + \eta_{k+1}.
\end{equation}

Substituting $x_{k+1}$ in the above equation results in,

\begin{equation}\label{output_withwatermark_eq_1}
    y_{k+1} = CAx_{k} + Cv_{k} + \eta_{k+1} + CB (u_{k - \Delta k}).
\end{equation}

Eqn.~\eqref{output_withwatermark_eq_1} establishes the relationship of input watermark to output, i.e., the sensor measurement. The last term in the above equation is the watermark signal generated randomly using the PLC clock. It shows that the output is a function of the watermarked control signal. This is the key result that analytically establishes the relationship between the input and output signals. \hfill $\blacksquare$ 

\noindent \textbf{Theorem 4.2} \emph{ Given the system model   in~\eqref{701}, Kalman filter gain $L$, and watermarked signal $\Delta k$, it can be shown that the difference between output signals is driven by the watermark signal and can be given as, $r_{k+1} =(CA - CLC)(x_k^a - \hat{x}_k^{wm})+ \eta_{k+1} + Cv_k    + CB(u_k^a - u_{k-\Delta k})$.}


\noindent \textit{\textbf{Proof}:} In the system model of eq.~\eqref{701}, attacker has access to the normal timing measurements and control signals and can replay those. Assuming that an adversary has access to the system model, and Kalman filter gain. During the replay attack using its knowledge an attacker tries to replay the data resulting in a system state described as,

\begin{equation}
    {x}_{k+1}^a = Ax_k^a +Bu_k^a + v_k 
\end{equation}
 and attacker's spoofed sensor measurements as,
 \begin{equation}
     {y}_{k+1}^a = C{x}_{k+1}^a + \eta_{k+1}
 \end{equation}
 
 
 \begin{equation}
           y_{k+1}^a = CAx_k^a + CBu_k^a + Cv_k + \eta_{k+1}
 \end{equation}

 However, using a watermarking signal in the control input $u_k$, the defender's state estimate becomes,
 \begin{equation}
     \hat{x}_{k+1}^{wm} = A\hat{x}_k^{wm} + Bu_{k - \Delta k} + L(y_k^a - C\hat{x}_k^{wm}),
 \end{equation}
where $\Delta k$ is the watermark signal. 

\begin{equation}
    \hat{y}_{k+1}^{wm} = C\hat{x}_{k+1}^{wm}
\end{equation}


\begin{equation}
    \hat{y}_{k+1}^{wm} = CA\hat{x}_k^{wm} + CBu_{k -\Delta k} + CL(Cx_k^a - C\hat{x}_k^{wm})
\end{equation}


The residual distribution between the watermarked and replayed control can be given as,

\begin{equation}
    \begin{split}
    y_{k+1}^a - \hat{y}_{k+1}^{wm} = CAx_k^a + CBu_k^a + Cv_k + \eta_{k+1} - CA\hat{x}_k^{wm} \\  - CBu_{k - \Delta k} - CLCx_k^a + CLC\hat{x}_k^{wm}
    \end{split}
\end{equation}


\begin{equation}
\label{watermark_proof}
    \begin{split}
    r_{k+1} = (CA - CLC)(x_k^a - \hat{x}_k^{wm})+ \eta_{k+1} + Cv_k  \\  + CB(u_k^a - u_{k-\Delta k}). 
    \end{split}
\end{equation}

\noindent There are four terms in the above equation. First-term is the difference between the attacker's spoofed state and the estimate of the system state for the watermarked signal. The second and third terms are sensor and process noise, respectively. The last term above is the difference in the control command execution times during a replay attack and watermark injected signal. The attacker's goal is to predict the delay $\Delta k$ to achieve $u_k^a = u_{k - \Delta k}$ so that the last term becomes zero and the watermark signal is countered. This is not possible since $\Delta k$ is randomly generated inside PLC at each time instance. \hfill $\blacksquare$

\section{Use Case: Power Generation System}
\label{sec_use_case_EPIC}

To validate the generalization of the proposed technique and applicability, a power generation subsystem is taken as a case study from an electric power smart grid testbed~\cite{kandasamy2019investigation_synchronization_attack}. This subsystem consists of two conventional generators (10 KVA each) which are run using 15kW variable speed drive motors implementing a combination of prime-mover and generator. Since our work focuses on the transient part of a process, therefore, we consider the synchronization process of the two generators. Synchronization takes place when a new generator starts working and it needs to be connected (to meet the demand) in parallel to the rest of the generators in the grid. Before putting the second generator online following conditions shall be met, a) The frequency of the generator must be the same as the frequency of the line/grid, b) The magnitude of the generator’s voltage must be the same as the magnitude of line/grid voltage, c) The phase angle of the generator’s voltage must be the same as the phase angle
of the line/grid voltage. The first two parameters do not depend on the state of the line/grid. However, the third parameter i.e., phase angle depends on the state of the phase angle of
the line/grid. The parallel connection is enabled by a circuit breaker which closes
once the phase angle difference is approximately equal to zero (usually around
$10^{\circ}$ in practical cases). We will measure the time it takes to get synchronized for each of these two generators. The time taken is the function of the device as the phase angle difference depends on the speed of the motor and can be given as,
\begin{equation}
    \frac{d\phi_d}{dt} = RPM_0 \times \Delta RPM,
\end{equation}

\begin{figure}
    \centering
    \includegraphics[height=2.5cm,width=9cm]{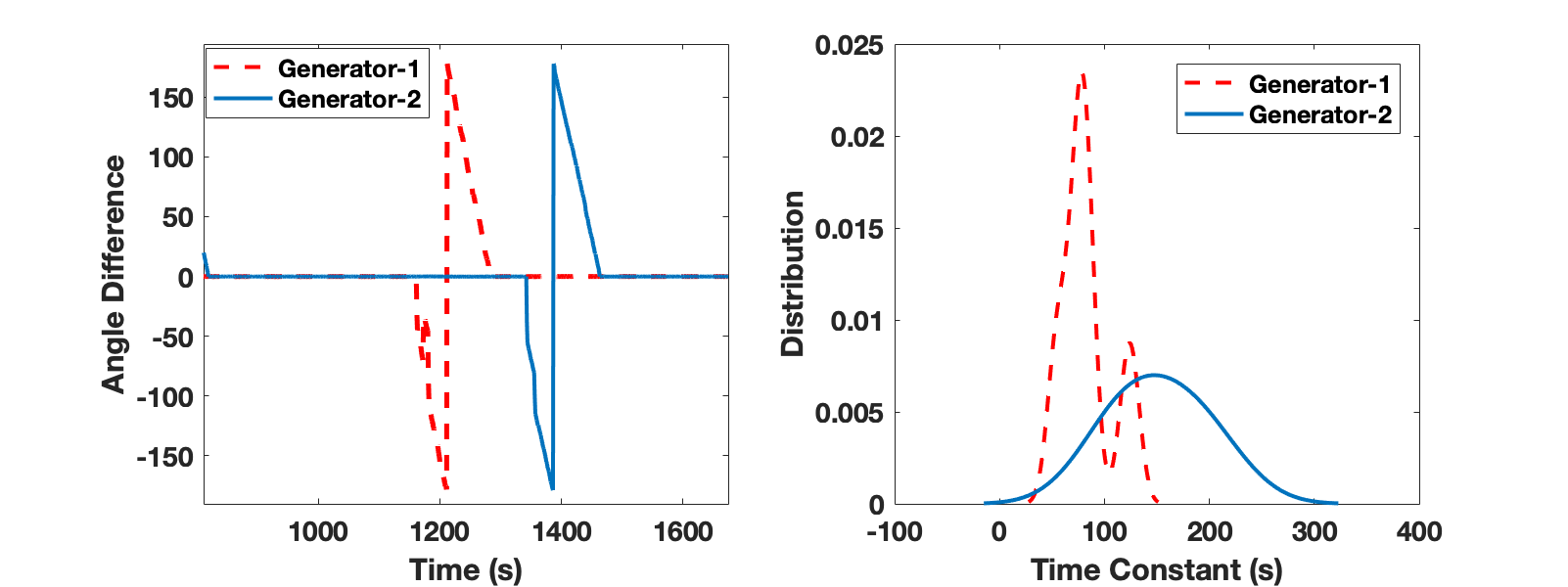}
    \caption{Synchronization process for two generators. It shows the angle difference between the two generators, during the synchronization process, the x-axis is the time to synchronize and reach the steady state (angle difference goes to zero).}
    \label{fig:SynchronizationProcess}
\end{figure}

\noindent Where $\Delta RPM = RPM_0  - RPM$, $RPM_0$ is the reference speed and RPM is the actual speed. The change in $\phi_d$ is cyclical and varies between $180^{\circ}$ to $-180^{\circ}$. Figure~\ref{fig:SynchronizationProcess} shows the generator synchronization process. The top plot shows one instance for each of the generators during the synchronization process, the x-axis shows the time taken for synchronization in seconds and the y-axis shows the angle difference between the two generators. When synchronized, the angle difference goes to zero and the system achieves the steady-state. The bottom plot (Figure~\ref{fig:SynchronizationProcess}) shows the distribution for the time taken by each generator to synchronize referred to as \tc in this paper. It is evident that both generators can be distinguished based on \tc distributions.

\begin{figure}
    \centering
    \includegraphics[scale=0.23]{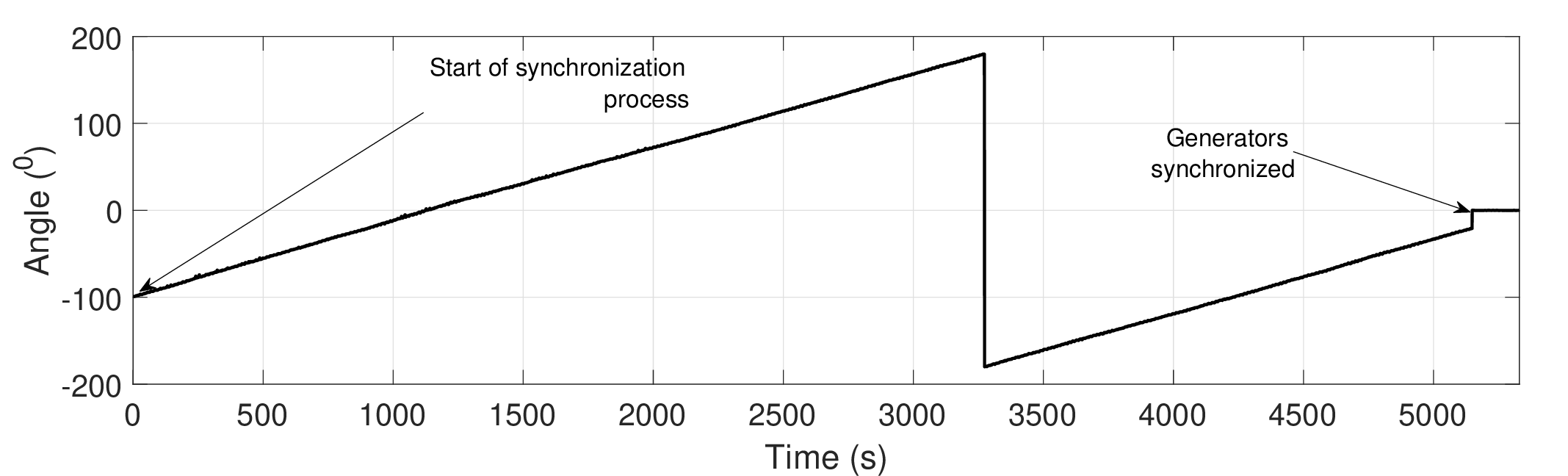}
    \caption{Experimental results of synchronization process of two generators after the launch of attack~\cite{kandasamy2019investigation_synchronization_attack}. The synchronization process that took time in seconds is delayed for 1.5 hours, leaving the generators unavailable.}\vspace{-3mm}
    \label{fig:experiment_attacked}
\end{figure}

The rationale for choosing a generator synchronization process is a recent work~\cite{kandasamy2019investigation_synchronization_attack} proposing practical DoS attacks on a smart grid by delaying the synchronization process indefinitely hence making the generators unavailable when needed. An attack on the synchronization process is launched to delay it from the order of seconds to several minutes. The details of the attack can be found in \cite{kandasamy2019investigation_synchronization_attack} but a summary is provided here. The normal transient process for the generator is to be completed in few seconds and make the generator available for meeting the power demand but this attack was able to delay the synchronization process by hours. Figure~\ref{fig:experiment_attacked} shows that the synchronization process is delayed by more than 1.5 hours and it deviates from the time constant of the generator hence getting detected. 


\section{Related Work}
\label{sec:Related_Work}
\noindent\emph{Transient-based Device Fingerprinting}: The ideas based on the transient fingerprints of a physical phenomenon have been discovered for wireless signals. RF fingerprints have been proposed using a measured temporal link signature to uniquely identify the link between a transmitter and a receiver\,\cite{hall2005,gerdes2006}. Researchers have shown that signal waveforms can also be used to extract the transient phase at the start of wireless transmission for device fingerprinting amid manufacturing inconsistencies\,\cite{hall2005}. Besides wireless fingerprinting,  it is possible to fingerprint devices based on transient temporal response in analog signals from Ethernet NICs\,\cite{gerdes2006}. However, the work closest to the proposed technique is based on physical operation time (opening/closing times) of a latching relay~\cite{raheem2016}. The analysis is carried out on $2$ latching relays based on their operation times. That work practically demonstrated that due to the construction of the relays it is possible to fingerprint the actuators as well as the states of the actuators, e.g., open/close. However, the limitation of that idea is that it extracts the timing features from the network layer traffic that itself is susceptible to modifications~\cite{john_acns2017_castellanos2017legacy,urbina_CCS2016limiting}, as discussed in the threat model in section ~\ref{sec:threat_model}. 

\noindent \emph{Challenge-Response/Watermarking based Authentication}: The idea of challenge-response protocols is well-known in the cyber security domain. Recently, there are few research efforts exploring techniques around the broader domain of challenge-response design~\cite{shoukry2015,bruno_mo_physicalwatermarking_2015,ahmed2020noisensePrint_ACMTOPS}. Physical Challenge-Response Authentication (PyCRA)~\cite{shoukry2015} proposed a defense technique for the case of active sensors, i.e., those with a transmitter and a receiver, e.g., an ultrasonic sensor that transmits an ultrasonic signal and measures the distance based on received signal reflected from an obstacle. This is achieved by randomly turning the active transmitter ON/OFF to figure out the presence of an attacker. However, it has been shown~\cite{sampling_race2016} that PyCRA fails if an attacker machine has a higher sampling rate than the defender and gets to learn the ON/OFF pattern. Another proposal is to add the challenge from the physical process side~\cite{ahmed2020noisensePrint_ACMTOPS}, however, this would be challenging in many practical applications, e.g., adding a challenge from inside a water pipe or on a high voltage transmission line. Moreover, both of these techniques discussed above, focus on sensor authentication. One of the earlier study~\cite{bruno_mo_physicalwatermarking_2015} on watermarking the control signal is closely related to our work. Mo et al.~\cite{bruno_mo_physicalwatermarking_2015} proposed to modify the optimal control signal with a noise component (called watermark) to detect the replay attacks. As highlighted within that study adding noise in the control signal results in sub-optimal control which is not desired in the real-world application. The contributions from Mo's work on a simulated scenario are interesting but lack the novel design that would be practical for a real-world system. We proposed a design for a watermarking technique that is built on top of device and process timing response, rather than modifying the control itself, we propose to add a random delay in the optimal control signal. Moreover, we have tested the proposed watermark design on the real water plant in contrast to earlier simulation-based works. 
We believe this is the first work that is based on our novel idea of actuator and process fingerprinting and comprehensively carried out an experimental study on a real testbed.


\section{Supporting Tables and Figures}
\label{appendix_supporting_table}

\begin{figure}[t]
    \centering
    \includegraphics[scale=0.035]{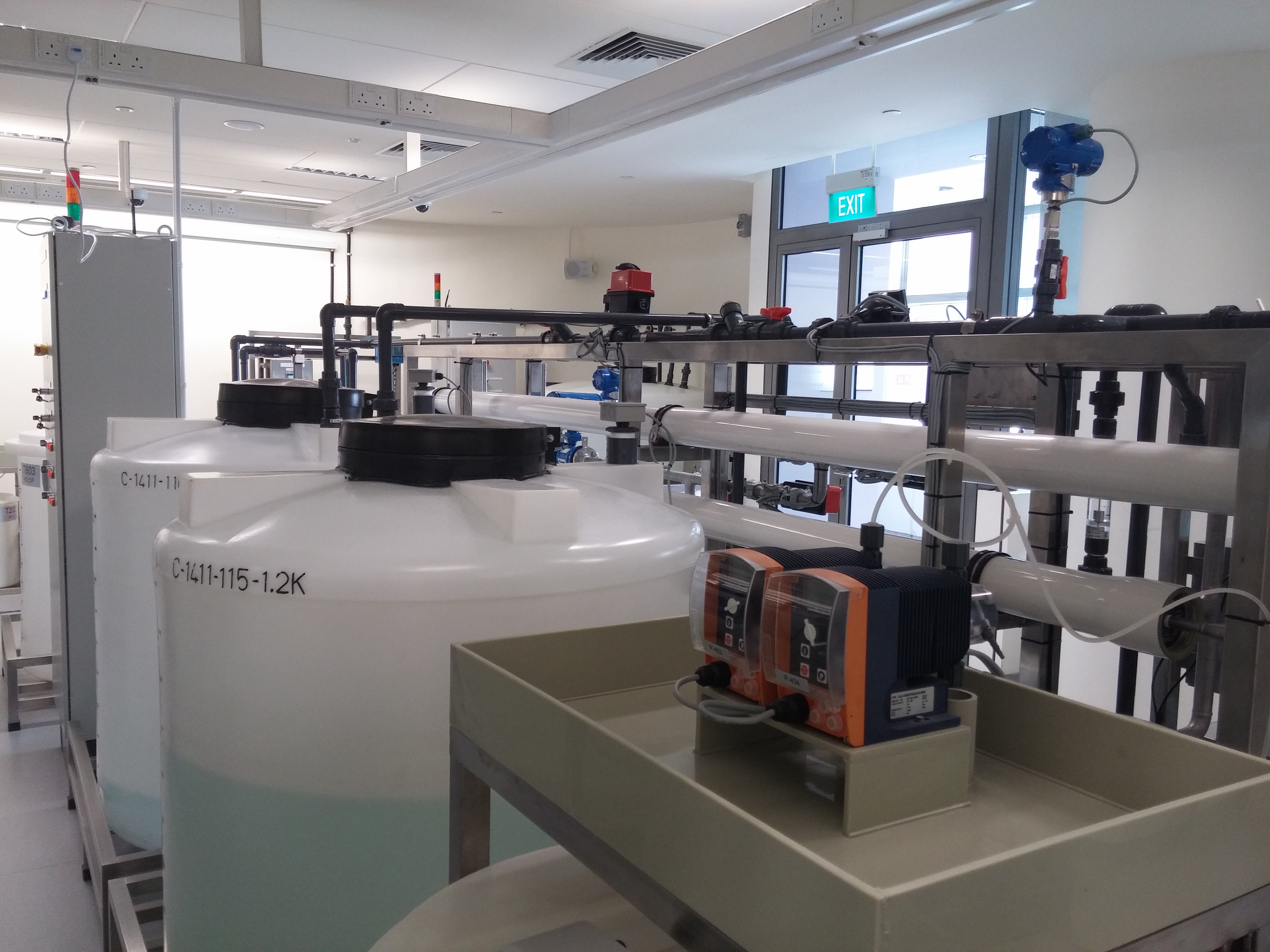}
    \caption{SWaT Testbed used in this study.} 
    \label{fig:swat_pic}
\end{figure}

\begin{figure}
    \centering
    \includegraphics[scale=0.17]{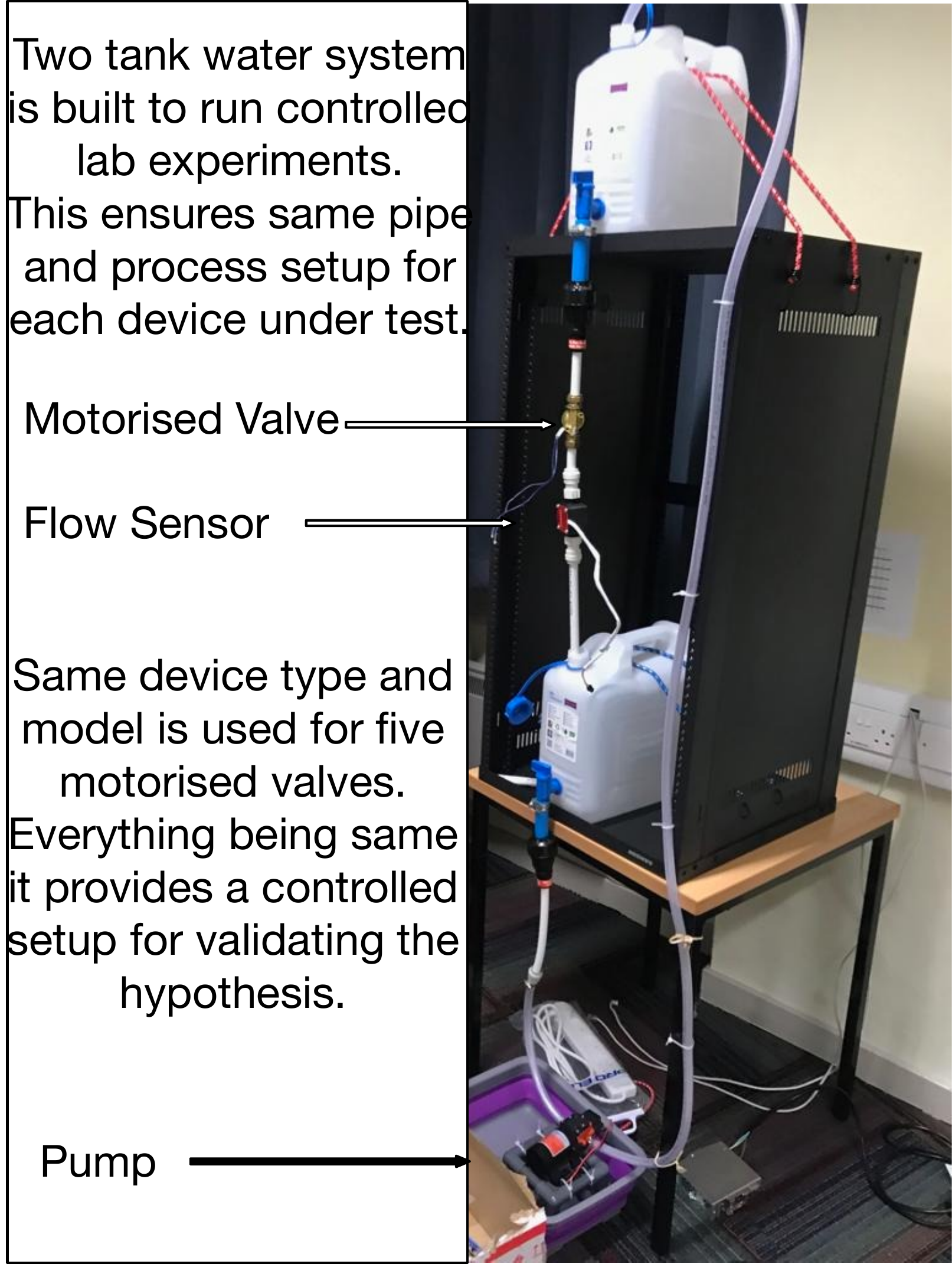}
    \caption{Two-tank water loop testbed for 5 motorised valves.}
    \label{fig:water_loop_testbed}
\end{figure}

\begin{figure}[htp]
    \centering
    \begin{adjustbox}{max width=0.4\textwidth}
        \includegraphics[width=0.4\textwidth]{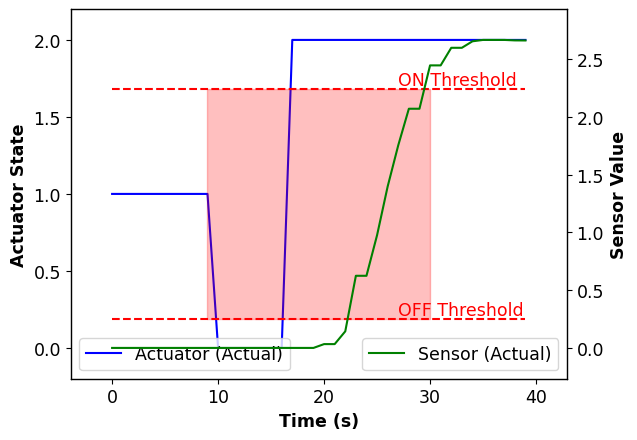}
    \end{adjustbox}
    \caption{Transition time of an actuator's ON operation as measured by our CUSUM-based detector. The transition time is represented by the horizontal length of the area highlighted in red, and it starts from the moment when the actuator changes state from ON/OFF, and ends at the moment when the associated sensor first reaches the threshold.}
    \label{fig:detector_detection_threshold}
\end{figure}

Table \ref{features} provides a detailed list of the features utilized in our experiments. Figure \ref{fig:swat_pic} illustrates the SWaT Testbed employed for our research, while Figure \ref{fig:water_loop_testbed} depicts the two-tank water loop testbed designed for five motorized valves. Additionally, Figure \ref{fig:detector_detection_threshold} demonstrates the transition time of an actuator’s ON operation, as measured by our CUSUM-based detector. 






\begin{table}[htp]
\begin{center}
\caption{List of features used.  }
\label{features}
 \begin{tabular}[!htb]{|l | l|} 
 \hline
 {\bf Feature} & {\bf Description}  \\ 
 \hline
 Mean & $\bar{x} = \frac{1}{N} \sum_{i=1}^{N} x_i$  \\ 
 \hline
 Std-Dev & $\sigma = \sqrt[]{\frac{1}{N-1}\sum_{i=1}^{N}(x_i - \bar{x}_i)^2}$  \\
 \hline
 Mean Avg. Dev & $D_{\bar{x}} = \frac{1}{N}\sum_{i=1}^{N} |x_i - \bar{x}|$  \\
 \hline
 Skewness & $\gamma = \frac{1}{N}\sum_{i=1}^{N}(\frac{x_i - \bar{x}}{\sigma})^3$  \\
 \hline
 Kurtosis & $\beta = \frac{1}{N} \sum_{i=1}^{N}(\frac{x_i - \bar{x}}{\sigma})^4 - 3$  \\
 \hline
 Spec. Std-Dev & $\sigma_s = \sqrt[]{\frac{\sum_{i=1}^{N}(y_f(i)^2)*y_m(i)}{\sum_{i=1}^{N}y_m(i)}}$  \\
 \hline
 Spec. Centroid & $C_s = \frac{\sum_{i=1}^{N}(y_f(i))*y_m(i)}{\sum_{i=1}^{N}y_m(i)}$  \\
 \hline
 DC Component & $y_m(0)$  \\ [1ex] 
 \hline
\end{tabular}
\end{center}
\end{table}

\end{document}